\documentclass[prd,twocolumn,aps,superscriptaddress,amsmath,amssymb,amsfonts,nofootinbib]{revtex4-2}

\pdfoutput=1

\usepackage{graphicx}
\usepackage{xcolor}
\usepackage{hyperref}
\usepackage{amsmath}
\usepackage{slashed}
\usepackage[normalem]{ulem}
\usepackage[capitalise]{cleveref}
\RequirePackage{xspace}
\newcommand{\etal}{\emph{et al.}\xspace}
\newcommand{\UNIT}[1]{\ensuremath{\,{\rm #1}}\xspace}
\newcommand{\tr}{\mathrm{tr}}
\newcommand{\iu}{\ensuremath{\mathrm{i}}}

\newcommand{\MeV}{\UNIT{Me\kern-0.15ex V}}
\newcommand{\GeV}{\UNIT{Ge\kern-0.15ex V}}
\newcommand{\fm}{\UNIT{fm}}

\begin{document}

\newcommand{\UCY}{Department of Physics, University of Cyprus, P.O. Box 20537, 1678 Nicosia, Cyprus}
\newcommand{\CASTORC}{Computation-based Science and Technology Research Center, The Cyprus Institute, 20 Kavafi Str., Nicosia 2121, Cyprus}
\newcommand{\Temple}{Department of Physics, Temple University, 1925 N. 12th Street, Philadelphia, PA 9122-1801, USA}
\newcommand{\TUB}{Technical University of Berlin, Berlin, Germany}

\title{The nucleon unpolarized generalized form factors and Mellin moments up to fourth order}

\author{Constantia Alexandrou}\affiliation{\UCY}\affiliation{\CASTORC}
\author{Simone Bacchio}\affiliation{\CASTORC}
\author{Martha Constantinou}\affiliation{\Temple}
\author{Christian Kummer}\affiliation{\UCY}\affiliation{\TUB}
\author{Yan Li}\affiliation{\CASTORC}
\author{Gregoris Spanoudes}\affiliation{\UCY}

\begin{abstract}
Nucleon Mellin moments of parton distribution  are computed up to  the fourth order  in lattice QCD.  The computation is performed using one ensemble of twisted mass fermions at the physical pion mass point. We employ boosted frames  to access the higher-order Mellin moments of generalized parton distributions.  We also extract the forward-limit Mellin moments $\langle x^{n-1}\rangle$  for $n=2,3,4$. These Mellin moments are used to construct  unpolarized  parton distribution functions and compare to phenomenological extractions.   
\end{abstract}

\maketitle
\section{Introduction}
One of the most important laboratories for exploring non-perturbative QCD is the proton. As a stable and abundant particle, the proton serves as a natural probe of strong-interaction dynamics. High-precision experiments are being conducted across a spectrum of energies at world-leading facilities, such as MAMI in Mainz, GSI in Darmstadt, Jefferson Lab in Virginia, and Fermilab in Illinois. At higher energies, the Large Hadron Collider (LHC) at CERN provides complementary insights. Many of these experiments aim not only to elucidate the proton structure but also to test the Standard Model to unprecedented accuracy, potentially revealing signatures of new physics. However, extracting meaningful results from such experiments necessitates a precise theoretical understanding of QCD effects, particularly in the low-energy, non-perturbative regime.

A powerful theoretical framework for describing the internal structure of hadrons is provided by the generalized parton distributions (GPDs). GPDs encode rich information about the spatial and momentum distribution of quarks inside the nucleon. They generalize the concept of parton distribution functions (PDFs) by including information about the transverse position and spin correlations of partons. While PDFs can  be interpreted probabilistically, GPDs are more subtle; in general, they describe interference effects between quantum amplitudes. In a certain kinematic limit, such as in the impact-parameter representation, they can be interpreted as spatial distributions of quarks~\cite{Burkardt:2000za}. For a comprehensive overview of GPDs, we refer  to Ref.~\cite{Diehl:2003ny}.

In this study, we focus on the moments of the unpolarized GPDs, denoted by H and E, which are defined through the decomposition of the nucleon matrix element of the non-local operator:
\begin{widetext}
\begin{align}
     \int \frac{{\rm d}\lambda}{2\pi} \, e^{\iu \lambda x} \,
     \langle N(p^\prime) |\overline{\psi}(-\lambda n/2) \gamma^\mu \mathcal{W}(-\lambda n/2,\lambda n/2) \psi(\lambda n/2)| N(p) \rangle 
     = \overline{u}(p^\prime) \bigl[ \gamma^\mu H(x,\xi,q^2) + \frac{i\sigma^{\mu\nu}q_\nu}{2m_N} E(x,\xi,q^2) \bigr] u(p) \,,
     \label{eq:def GPDs}
\end{align}
\end{widetext} where $n$ is a light cone vector and the quark fields $\psi$ are connected by the Wilson-line $\mathcal{W}(-\lambda n/2,\lambda n/2)$ in a gauge-invariant manner. These GPDs are not directly observable; instead, they appear in convolution integrals in scattering amplitudes and must be extracted from global analyses of experimental data~\cite{Hagler:2009ni,Diehl:2003ny}. Their significance is underscored by the fact that the Electron-Ion Collider (EIC), a major upcoming facility in the United States, is specifically designed to explore nuclear structure and determine GPDs with high precision.

In this work, we compute  the Mellin moments of the unpolarized GPDs or the generalized form factors (GFFs). From a theoretical perspective they are central because they connect GPDs to local operator matrix elements, which are directly calculable in lattice QCD without the need for inverse transformations~\cite{Ji:1998pc}. Furthermore, they serve as inputs to sum rules, for instance, Ji’s sum rule connects the second moment of GPDs to the total angular momentum carried by quarks~\cite{Ji:1996ek}. Their importance does not stop there, they can also form a bridge between elastic and inelastic processes, as they interpolate between form factors and moments of PDFs~\cite{Ji:1996ek,Radyushkin:1997ki}. Through the operator product expansion they provide a systematic way to study the QCD evolution of GPDs~\cite{Diehl:2003ny}. 

Despite their relevance, higher Mellin moments, especially beyond the first or second, remain underexplored in lattice QCD. Most studies to date rely on early lattice ensembles with unphysical parameters. For example, until recently the only lattice calculation of nucleon matrix elements involving two-derivative operators dates back to 2005~\cite{Gockeler:2005vw}, and used ensembles with pion masses around 600 MeV, far from the physical pion mass point. With advances in lattice methodology, computing power, and the availability of near-physical pion mass gauge configurations, there is a pressing need to revisit and improve these results. Recently, there have been advancements in the calculation of the pion Mellin moments up to sixth order~\cite{Francis:2025pgf,Francis:2025rya,Detmold:2025lyb}, but they only consider the forward limit and used ensembles with pion mass heavier than physical. 

This study aims to address this gap by performing lattice calculations of GPD moments, focusing particularly on higher-order GFFs that are both phenomenologically valuable and theoretically illuminating. These moments are of growing interest to the experimental community, as they provide robust benchmarks for global fits and help constrain the models used in interpreting data from current upcoming facilities like the EIC. By improving the accuracy of Mellin moment and GFF calculations, we enhance the theoretical foundation necessary for interpreting future experimental results and deepen our understanding of nucleon structure from  QCD. 

\section{Generalized Form Factors}
\subsection{Definitions}
The twist-two vector operators are defined by
\begin{align}
    \mathcal{O}^{\{\mu_1 \mu_2 ...\mu_n\}} = \overline{\psi} \, \gamma^{\{\mu_1} i \overleftrightarrow{D}{}^{\mu_2} ... i\overleftrightarrow{D} {}^{\mu_n \}}\,\psi\,,
    \label{eq:operator}
\end{align} where the flavor indices have been suppressed and the curly brackets denote symmetrization of the indices and removal of traces. The symmetric derivative
$
    \overleftrightarrow{D_\mu} = \frac{1}{2}(\overrightarrow{D_\mu} - \overleftarrow{D_\mu})
$
is defined by the difference between forwards- and backwards-acting derivatives. These leading twist operators appear, as an example, in the operator product expansion in DIS~\cite{Itzykson:1980rh}. Since they are not related to any QCD symmetry, their matrix elements offer valuable insights to dynamical information on nucleon structure. The matrix element can be written in Minkowski space as~\cite{Hagler:2003jd}
\begin{align}
    \langle N(p^\prime)& | \mathcal{O}^{\{\mu_1 \mu_2...\mu_n \}} | N(p) \rangle = \nonumber \\
    \overline{u}(p^\prime) \bigg[& \sum_{\substack{i=0\\{\rm even}}}^{n-1} \Bigl\{ q^{\{\mu_1}...q^{\mu_i} P^{\mu_{i+1}}...P^{\mu_{n-1}} \gamma^{\mu_n \}} A_{n,i}(q^2) \nonumber \\
    &+q^{\{\mu_1}...q^{\mu_i} P^{\mu_{i+1}}...P^{\mu_{n-1}} \frac{i \sigma^{\mu_n \} \alpha}q_\alpha}{2m_N} B_{n,i}(q^2) \Bigr\} \nonumber \\
    &+\frac{q^{\{\mu_1}q^{\mu_2}...q^{\mu_n \}}}{m_N} C_{n,0}(q^2) \Big|_{n\,{\rm even}} \bigg] u(p) \,,
    \label{eq:def gff}
\end{align} where $\bar{u}(p^\prime)$, $u(p)$ are nucleon spinors,  and $P^\mu=(p^{\prime \mu}+p^\mu)/2$ and $q^\mu=p^{\prime\mu}-p^\mu$ are the average and the difference between the initial and final nucleon four-momenta, respectively. The generalized form factors are defined as the coefficients $A_{n,i}$, $B_{n,i}$ and $C_{n,0}$ of \cref{eq:def gff}. They are, in fact, well-defined, as \cref{eq:def gff} contains all Lorentz-invariant contributions consistent with parity and time invariance~\cite{Ji:1998pc,Diehl:2003ny}.

The GFFs are related to the moments of the GPDs $H(x,\xi,q^2)$ and $E(x,\xi,q^2)$ by
\begin{align}
    \int_{-1}^1 {\rm d}x\, &x^{n-1} H(x,\xi,q^2) \nonumber\\
    &=  \sum_{\substack{i=0\\{\rm even}}}^{n-1} A_{n,i}(q^2)(-2\xi)^i + C_{n0}(q^2)(-2\xi)^n \Big|_{n\,{\rm even}} \,,\nonumber\\
    \int_{-1}^1 {\rm d}x\, &x^{n-1} E(x,\xi,q^2) \nonumber \\
    &=  \sum_{\substack{i=0\\{\rm even}}}^{n-1} B_{n,i}(q^2)(-2\xi)^i - C_{n0}(q^2)(-2\xi)^n \Big|_{n\,{\rm even}} \,,
\end{align} with the Bjorken-$x$, the fraction of the nucleon momentum that is being carried by a quark. Note how the moments are polynomial of even powers in the skewness $\xi$. In particular, in the forward limit, the $n^{\rm th}$ Mellin moment is given by 
\begin{align}
    \int_{-1}^1 {\rm d}x\, x^{n-1} &H(x,0,0) \nonumber \\
    &= \int_{-1}^1 {\rm d}x\, x^{n-1} q(x) 
    =\langle x^{n-1} \rangle = A_{n0}(q^2=0) \,.
    \label{eq: momentsGFF}
\end{align} For completeness, the explicit decompositions of \cref{eq:def gff} for one, two and three derivatives in Minkowski space are
\begin{widetext}
    \begin{align}
         \langle N(p^\prime) | \mathcal{O}^{\{\mu \nu \}} | N(p) \rangle &= \overline{u}(p^\prime) \bigg[ P^{\{\mu}\gamma^{\nu\}} A_{20}(q^2) +  \frac{\iu  P^{\{\mu}\sigma^{\nu \} \alpha}q_\alpha}{2m_N} B_{20}(q^2) + \frac{q^{\{\mu}q^{\nu\}}}{m_N}+C_{20}(q^2) \bigg] u(p) \,, \nonumber \\
          \langle N(p^\prime) | \mathcal{O}^{\{\mu \nu \rho \}} | N(p) \rangle &= \overline{u}(p^\prime) \bigg[ P^{\{\mu}P^\nu \gamma^{\rho\}} A_{30}(q^2) +q^{\{\mu}q^\nu \gamma^{\rho\}} A_{32}(q^2) +  \frac{\iu  P^{\{\mu}P^\nu\sigma^{\rho \} \alpha}q_\alpha}{2m_N} B_{30}(q^2) +  \frac{\iu  q^{\{\mu}q^\nu\sigma^{\rho \} \alpha}q_\alpha}{2m_N} B_{32}(q^2) \bigg] u(p) \,, \nonumber\\
          \langle N(p^\prime) | \mathcal{O}^{\{\mu \nu \rho \lambda\}} | N(p) \rangle &= \overline{u}(p^\prime) \bigg[ P^{\{\mu}P^\nu P^\rho \gamma^{\lambda\}} A_{40}(q^2) +q^{\{\mu}q^\nu P^\rho \gamma^{\lambda\}} A_{42}(q^2)\nonumber \\
          +  &\frac{\iu  P^{\{\mu}P^\nu P^\rho\sigma^{\lambda \} \alpha}q_\alpha}{2m_N} B_{40}(q^2) +  \frac{\iu  q^{\{\mu}q^\nu P^\rho\sigma^{\lambda \} \alpha}q_\alpha}{2m_N} B_{42}(q^2) + \frac{q^{\{\mu}q^\nu q^\rho q^{\lambda\}}}{m_N} C_{40}(q^2) \bigg] u(p) \,.
          \label{eq:allgffs}
    \end{align}
\end{widetext} The kinematic factors in front of the GFFs are calculated numerically and require the nucleon mass $m_N$, as both $P$ and $q$ are dependent on it besides the mass appearing explicitly in these equations as well. 

\subsection{Operator Mixing\label{sec:Mixing}}
In continuum  Minkowski spacetime, the derivative operators of Eq. \eqref{eq:operator} are classified according to the irreducible representations of the Lorentz group $SO(1,3)$. After Wick rotation, the relevant symmetry group becomes the Euclidean rotation group $O(4)$. Due to the discreteness of the lattice, $O(4)$ breaks down to the hypercubic group $H(4)$~\cite{Beccarini:1995iv}. As a consequence, the operators entering \cref{eq:def gff} can mix on the lattice with other operators of the same or lower dimension that transform identically under $H(4)$. This typically leads to a larger mixing pattern than in the continuum, particularly for operators containing repeated Lorentz indices. Such enhanced mixing complicates both the renormalization procedure and the extraction of GFFs, as multiple matrix elements are involved. Mixing with lower-dimensional operators is especially problematic due to the presence of power divergences, which scale as $a^{-d}$, where $a$ is the lattice spacing and $d$ the difference in operator dimensions. The subtraction of these power-divergent contributions is delicate and must be accurate enough to ensure a reliable continuum limit. For this reason, operators exhibiting power-divergent mixing are generally avoided in practical calculations when possible.

Several strategies have been proposed to address operator mixing, each with its own merits and shortfalls. One approach is to determine the full renormalization matrix nonperturbatively by imposing suitable renormalization conditions and carefully accounting for all operators involved~\cite{Gockeler:1996mu}. While the procedure is conceptually straightforward, in practice it may lead to reduced statistical precision and can be more susceptible to systematic uncertainties, making it less robust than alternative methods.

Another straightforward approach is to restrict the analysis to operators that do not mix. This can be achieved by selecting operators with all Lorentz indices different. Such a condition can only be satisfied for operators with up to three derivatives, since operators with four or more derivatives necessarily involve more than four indices, implying that index repetition is unavoidable. In addition, operators with two or three derivatives and all distinct indices require nonzero spatial momentum, i.e. a boosted frame, for their matrix elements to be accessible. In the forward limit the GFFs $A_{n0}(0)$ are proportional to components of the  boost momentum, cf. \cref{eq:allgffs}. One index can be in the time-direction and thus  brings in the energy, but for two derivatives there must  be at least two non-zero spatial components of the boost momentum. Consequently one must boost the sink along two directions to obtain $\langle x^2 \rangle$. A boosted frame increases the statistical error and enhances cutoff effects~\cite{Beccarini:1995iv}.

In this work, we use the following operators: For the two-derivative operator we use $\mathcal{O}^{\{\mu\nu\rho\}}$ with $\mu\neq\nu\neq\rho\neq\mu$. Since all indices are different from each other, this operator does not mix. The ideal three-derivative operator to avoid mixing would be $\mathcal{O}^{\{1234\}}$. Unfortunately, this operator turns out to be  more noisy than other traceless operators, such as $\mathcal{O}^{\{44ij\}}$ ($i \neq j$) and  $\mathcal{O}^{\{44ij\}}-\mathcal{O}^{\{kkij\}}$ with $i \neq j \neq k \neq i$. There are two reasons for the larger noise: Firstly, the nucleon matrix element of $\mathcal{O}^{\{1234\}}$ is proportional to the nucleon momentum cubed, while $\mathcal{O}^{\{44ij\}}$ trades one factor of the momentum with the energy of the nucleon. In other words, the kinematic factor  of the matrix element is smaller for the operator with all indices different and thus the signal-to-noise is worse. Secondly, there are less equations, i.e. less data, to extract the GFFs for this operator: One can only permute the momentum directions. A non-fully symmetric operator under spatial rotations such as $\mathcal{O}^{\{44ij\}}$ on the other hand can give rise to  more equations. Therefore, we consider in our analysis two different traceless combinations $\mathcal{O}^{\{44ij\}}$ and $\mathcal{O}^{\{44ij\}}-\mathcal{O}^{\{kkij\}}$ with $i \neq j \neq k \neq i$, which are less noisy than $\mathcal{O}^{\{1234\}}$. These two operators exhibit mixing with other operators of the same dimension~\cite{Gockeler:1996mu}. Since the mixing is not power-divergent, we expect that these operators reproduce the three-derivative GFFs sufficiently well. Indeed, we found these two operators to be less noisy and in agreement with the operator $\mathcal{O}^{\{1234\}}$.  

Recently, it has been proposed to employ Wilson-flow to avoid operator mixing~\cite{Francis:2025pgf,Francis:2025rya}. By applying Wilson flow in both gauge and fermion fields of the operators, ultraviolet divergences are suppressed exponentially and their matrix elements become finite up to a multiplicative renormalization factor of the fermion field.  Then the continuum limit can be taken at fixed flow times, where rotational symmetry is restored. Using the short-flow time expansion~\cite{Luscher:2013vga}, perturbative matching at zero flow time can be performed in the continuum. Therefore, one is not limited to operators with at most three derivatives, and higher-order derivatives become calculable. In particular, the operator $\mathcal{O}^{\{ 4 4 ...4 \}}$ can be used, which can be evaluated in the lab frame reducing statistical noise. However, this method has only been used for the pion with larger than physical pion masses and further testing is needed before doing a computation of the nucleon at physical pion mass. Furthermore, the approach has been applied in the forward limit where the contractions, which  must be done at many values of the flow times,  can be optimized. When there is momentum transfer, as considered in this work, such optimization is not possible and the computation becomes much more expensive.

\section{Methodology\label{sec:Procedure}}
\subsection{Gauge Ensembles}
We use one gauge ensemble generated with $N_f = 2+1+1$ dynamical quark flavors on a $64^3 \times128$ lattice volume by the European Twisted Mass Collaboration (ETMC)~\cite{Alexandrou:2018egz} with the twisted mass formulation~\cite{Frezzotti:2003ni,Frezzotti:2000nk} and including a clover term~\cite{Sheikholeslami:1985ij} that stabilizes calculations at the physical pion mass point. The  twisted mass fermion discretization scheme leads to an automatic $\mathcal{O}(a)$ improvement. The  degenerate light quark, strange and charm quark masses are tuned to reproduce approximately their physical values. The lattice spacing $a$ is set by demanding that the observed nucleon mass coincide with the physical nucleon mass~\cite{Alexandrou:2018sjm}. The most relevant parameters are given in \cref{tab:gauge-ensemble}, for more details we refer to~ Refs.\cite{Alexandrou:2018egz,Alexandrou:2018sjm}.

\begin{table}[h!]
    \centering
    \begin{tabular}{c c c c c c}
    \hline \hline
    Ensemble & $a$ [\fm] & $(L/a)^3\times T/a$ & $m_\pi$ [\GeV] & $L$ [\fm]  \\ 
    \hline
    cB211.072.64 & 0.0801(4) & $64^3\times 128$ & 0.1393(7) & 5.12(3) \\
    \hline \hline
  \end{tabular}
  \caption{Simulation parameters of the gauge ensemble used in this study.}
  \label{tab:gauge-ensemble}
\end{table}

\subsection{Correlation Functions}
In order to determine the matrix element of \cref{eq:operator}, we build the ratio of two- and three-point correlation functions  to cancel time dependent exponentials and unphysical overlaps. The nucleon two-point function is given by
\begin{align}
    &C(\Gamma_0, \vec{p}, t_s , t_0) =\sum_{\vec{x}_s} e^{-\iu (\vec{x}_s-\vec{x}_0)\cdot \vec{p}} \times \nonumber\\ &\tr[\Gamma_0 \langle J_N(t_s,\vec{x}_s) \bar{J}_N(t_0,\vec{x}_0)\rangle] \,,
    \label{eq:deftwopcorr}
\end{align} with the standard nucleon interpolator
\begin{align}
    J_N(t,\vec{x})= \epsilon^{abc} u^a(x) [u^{Tb}(x) \mathcal{C} \gamma_5 d^c(x)]\,,
\end{align} where $u$ and $d$ are the up- and down-quark spinors, $\mathcal{C}=\gamma_0\gamma_2$ is the charge conjugation matrix, and $\Gamma_0 = \frac12 (1+\gamma_0)$ is the unpolarized projector. The initial coordinate $x_0$ is referred to as the source and $x_s$ as the sink. We use Gaussian smeared quark fields in the nucleon interpolator as described in~\cite{Alexandrou:1992ti,Gusken:1989qx} with APE smeared gauge links~\cite{APE:1987ehd}. The smearing  increases the overlap of the nucleon interpolator with the ground state,  and reduces overlaps with excited states. For the APE smearing we apply 50 iterations with $\alpha_{\rm APE}=0.5$, and 95 steps with $\alpha_{\rm Gauss}=1.0$ for the Gaussian smearing.

The  three-point function is given by
\begin{align}
    C^{\{ \mu_1\mu_2...\mu_n\}}&(\Gamma, \vec{q}, \vec{p}^\prime, t_s, t_{\rm ins} , t_0) = \nonumber \\
    &\sum_{\vec{x}_{\rm ins},\vec{x}_s} e^{-\iu (\vec{x}_{\rm ins}-\vec{x}_0)\cdot \vec{q}} e^{-\iu (\vec{x}_s-\vec{x}_0)\cdot \vec{p}^\prime} \times \nonumber\\ \tr[\Gamma \langle J_N(t_s,\vec{x}_s) &\mathcal{O}^{\{ \mu_1\mu_2...\mu_n\}}(t_{\rm ins}, \vec{x}_{\rm ins}) \bar{J}_N(t_0,\vec{x}_0)\rangle] \,,
    \label{eq:defthreecorr}
\end{align} with sink momentum $\vec{p}^\prime$ and momentum transfer $\vec{q}=\vec{p}^\prime-\vec{p}$. The operator $\mathcal{O}^{\{ \mu_1\mu_2...\mu_n\}}(t_{\rm ins},\vec{x}_{\rm ins})$, is defined in \cref{eq:operator}. The three-point function is calculated either with the unpolarized projector $\Gamma_0$ or a polarized projector $\Gamma_k=\iu \gamma_5\gamma_k\Gamma_0$ with $k\in \{1,2,3\}$. The coordinate $x_{\rm ins}$ refers to the operator insertion. Without loss of generality the sink coordinate $x_0$ can be set to 0 in both \cref{eq:deftwopcorr} and \cref{eq:defthreecorr} since they only depend on the separation $x_s-x_0$, so from now on the dependence on the sink position will be dropped.

In this work, we only consider the connected three-point functions. For the isovector one naturally has a full cancellation of disconnected contributions to ${\cal{O}}(a^2)$. For the isoscalar, there are  disconnected contributions, which for the second derivative are about 25 \% compared to  the connected for this ensemble~\cite{Alexandrou:2020sml,Alexandrou:2017oeh,Yansnewpaper}. However, since in this work we are computing higher derivatives, we expect the disconnected contributions to be small since these become suppressed as the order of the moment increases. Therefore, we will neglect disconnected contributions in this work.

\subsection{Implementation of derivative operators}
The symmetric derivative, $\overleftrightarrow{D_\mu}$, has a forward- and a backward-acting derivatives
\begin{align}
    \overrightarrow{D_\mu} \psi(x) &= \frac{1}{2a}(U_\mu(x) \psi(x+\hat{\mu}) - U_{-\mu}(x)\psi(x-\hat{\mu}))\,,\nonumber\\
    \overline{\psi}\,\overleftarrow{D_\mu} &= \frac{1}{2a}(\overline{\psi}(x+\hat{\mu})U_{-\mu}(x+\hat{\mu}) - \overline{\psi}(x-\hat{\mu})U_\mu(x-\hat{\mu}))\,,
    \label{eq:def derivs}
\end{align} where the gauge-links $U_\mu$ ensure gauge-invariance and $\psi$ are the quark fields. Applying multiple such derivatives shifts the quark field again, which makes the resulting terms  more complicated but straight forward to compute, see for example Ref.~\cite{Beccarini:1995iv} for an explicit expression for the one- and two-derivative operators.

\subsection{Extraction of Matrix Elements\label{sec:extractionmatrixelements}}
In order to extract the matrix element of the operators given in \cref{eq:operator}, one  constructs ratios of three- and two-point functions. As both correlators decay exponentially, it is highly desirable to have the shortest possible source-sink separation in the two-point functions but still cancel unknown overlaps and exponential dependencies asymptotically. The ratio
\begin{align}
    &R^{\{ \mu_1 \mu_2...\mu_n\} }(\Gamma,\vec{q},\vec{p}^\prime,t_s,t_{\rm ins}) \nonumber\\
    &= \frac{C^{\{ \mu_1 \mu_2...\mu_n\} }(\Gamma,\vec{q},\vec{p}^\prime,t_s,t_{\rm ins})}{C(\Gamma_0,\vec{p}^\prime,t_s)} \times \nonumber\\
    &\sqrt{\frac{C(\Gamma_0,\vec{p},t_s-t_{\rm ins})C(\Gamma_0,\vec{p}^\prime,t_{\rm ins})C(\Gamma_0,\vec{p}^\prime,t_s)}{C(\Gamma_0,\vec{p}^\prime,t_s-t_{\rm ins})C(\Gamma_0,\vec{p},t_{\rm ins})C(\Gamma_0,\vec{p},t_s)}} \,,
    \label{eq:Ratio}
\end{align}  has these properties~\cite{Alexandrou:2013joa,Alexandrou:2011db,Alexandrou:2006ru}. Furthermore, in the ratio we use the two-point function with the same source positions as the three-point function for the given $t_s$, making use of correlations.

The matrix element of the operator \cref{eq:operator} is a linear combination of the $n$ GFFs, where $n=\#$derivatives +1, cf. \cref{eq:allgffs}. Now the GFFs only depend on $q^2$, while the matrix element depends on $\vec{p}^\prime$, $\vec{q}$, projector $\Gamma$ and the particular choice of indices. Thus, for a given $q^2$, there are $m$ unique matrix elements and in general $m\gg n$. Therefore, one has an over-constrained system of equations, and there are various ways to deal with it. One option is to first extract all $m$ matrix elements, and then solve the linear system of equations (LSE) for the GFFs. This would require determining $m+n$ parameters, which might lead to a larger error, i.e. the GFFs not representing the data well~\cite{Bali:2018zgl}. It is both more economical and less error-prone to first solve the LSE and then extract the matrix elements~\cite{Bali:2018zgl,Alexandrou:2019ali}. Concretely, we want to solve the LSE
\begin{align}
    R=MF\,,
    \label{eq: LSE}
\end{align} where $F$ is the vector of all GFFs of the considered derivative operator, $M$ is a matrix of the decompositions given in \cref{eq:allgffs} and $R$ are the ratios that have non-zero contribution. Note that the kinematic factors of the matrix element in general depend on the nucleon energy $p^0=E=\sqrt{m_N^2+\vec{p}^2}$. In order to obtain the nucleon mass, $m_N$, the effective energy $E_0$ is extracted by fitting the effective energy given by
\begin{align}
    aE_0 + \log\bigg( \frac{1+c_1 e^{-\Delta E\,t}}{1+c_1 e^{-\Delta E\,(t+1)}} \bigg) = \log\bigg(\frac{C(t)}{C(t+1)}\bigg)\,,
    \label{eq:eff mass}
\end{align} 
constructed using the two-point function at full statistics. 

The best linear unbiased estimator of $F$ in \cref{eq: LSE} is given by
\begin{align}
    F=(M^T {\rm cov(R)}M)^{-1} M^T {\rm cov}(R)R \,.
    \label{eq: WLS}
\end{align}
Ideally, one would use the full covariance matrix of $R$, but this renders the extraction unstable, so we use only its diagonal. This approach, known as weighted least squares, is numerically stable while still penalizing noisy data. Furthermore, it is not only consistent, but mathematically identical to the SVD normalized by the standard deviation of $R$ used in previous studies~\cite{Hagler:2003jd,Alexandrou:2019ali}.
By applying this method to all time separations $t_s$ and insertion times  $t_{\rm ins}$ the ratios $\mathcal{R}_{\rm GFF}(q^2,t_s,t_{\rm ins})$ corresponding to a given GFF are obtained. In the next step, the nucleon matrix element can be extracted. A final consideration, is whether \cref{eq: LSE} actually have a solution. In order to have a unique solution, there must be at least $n$ linearly independent equations, or in other words, $M$ must have full rank, i.e. rank($M$)=$n$. Unfortunately, this is not always the case, especially for matrix elements of the three derivative operators and  there are not enough equations at a particular $q^2$ to determine the five GFFs. One option is to assume that the GFFs are smooth enough to be roughly constant for small variations of $q^2$, and cluster close values of  $q^2$ together. This allows more $\vec{q}$ to enter into the LSE and, subsequently, more equations to be considered at once, so the we  can restore $M$ to full rank. This method has been attempted successfully for example in Ref.~\cite{Hackett:2023rif}.

\subsection{Treatment of excited states}
For large time separations, i.e. $t_s\Delta E \gg 1$, $(t_s-t_{\rm ins})\Delta E \gg 1$ where $\Delta E$ is the energy gap between the excited and ground state energies, the ratio $\mathcal{R}_{\rm GFF}(q^2,t_s,t_{\rm ins})$ converges to a time independent $\Pi_{\rm GFF}$. In practice, one is limited on how large $t_s$ can be since noise is increasing exponentially with increasing time separation. This means one must employ methods that can extract the ground state matrix element  at the smallest possible $t_s$.

The simplest such method is the so-called plateau method where one considers  that the ground  state dominates  and that excited states effects can be neglected. Thus, one fits the ratio $\mathcal{R}_{\rm GFF}$ to a constant. By fitting each $t_s$ separately one checks for convergence. This method can give very precise values since one fits only one parameter. A related method is fitting all the ratios with a $t_s \geq t_s^{\rm low}$, for a given $t_s^{\rm low}$. This is our preferred method of extracting the ground state, but one must verify that this method provides robust results by comparing with other methods and the ratio itself.

An alternative approach to extract $\Pi_{\rm GFF}$ is to use the summation method. In this technique, one also assumes ground state dominance but  sums the ratio $\mathcal{R}_{\rm GFF}(q^2,t_s,t_{\rm ins})$ over all insertion times $t_{\rm ins}$ not including the contact points on either side. This summation yields the quantity 
\begin{align}
    &S_{\rm GFF}(q^2,t_s)=\sum_{t_{\rm ins}=\tau}^{t_s-\tau} \mathcal{R}_{\rm GFF}(q^2,t_s,t_{\rm ins})  \nonumber \\
    =& c+\Pi_{\rm GFF}\times t_s + \mathcal{O}(e^{-\Delta E\, t_s})\,,
\end{align} 
where the ground state matrix element can be extracted as the slope of a linear fit. While  still being only a one-state fit like the plateau method, this method converges faster. It is roughly equivalent to the plateau method with a source-sink separation $t_s$ roughly twice as large~\cite{Alexandrou:2020sml}. On the other hand the summation method is more noisy than the plateau fit.

\subsection{Statistics}
In this study, three- and two- point functions are produced using different source positions per gauge configuration. 
\begin{table}[htb]
    \centering
    \begin{tabular}{c c c c c c}
    \hline \hline
    $\vec{n}^\prime$ & $N_{\rm mom}$ & $t_s/a$ & $N_{\rm conf}$ & $N_{\rm srcs}$  & $N_{\rm meas}$  \\ 
    \hline
    \multicolumn{5}{c}{two derivatives}\\
    $(1,1,0)$ & 12 & 8 & 735  & 1 & 8820 \\
    $(1,1,0)$ & 12 & 10 & 735  & 3 & 26460 \\
    $(1,1,0)$ & 12 & 12 & 735  & 9 & 79380 \\
    $(1,1,0)$ & 12 & 14 & 735  & 18 & 158760 \\
    \hline
    \multicolumn{6}{c}{three derivatives}\\
    $(1,1,1)$ & 8 & 8 & 735  & 1 & 5880 \\
    $(1,1,1)$ & 8 & 10 & 735  & 3 & 17640 \\
    $(1,1,0)$ & 12 & 14 & 735  & 18 & 158760 \\
    \hline \hline
  \end{tabular}
  \caption{The total statistics of the three- and two-point functions used in  the ratio of \cref{eq:Ratio}. The first column indicates a representation of the sink momentum $\vec{n}$, the second the number of momenta $N_{\rm mom}$ with the same $\vec{n}^2$,  and the third gives the source-sink separations. The number of source positions for each time separation is $N_{\rm srcs}$ and $N_{\rm conf}$ is the number of gauge configurations, which is the same for all of them. The total number of measurements is given by $N_{\rm meas}=N_{\rm mom}\times N_{\rm conf}\times N_{\rm srcs}$.}
  \label{tab:statistics}
\end{table}
For each of these three-point functions, we  use two-point functions using the same source positions so that we make maximum usage of the correlations between numerator and denominator in the ratio of \cref{eq:Ratio}. In order for us to determine the two- (three-) derivative operators without mixing we need the sink to be boosted at least along two (three) directions. The smallest such boosts are $(1,1,0)$ and $(1,1,1)$, which have $\vec{n}^2=2$ and $\vec{n}^2=3$, respectively. We also calculate all possible permutations of the components of the sink momentum and all possible sign flips to get the number of possible sink momenta $N_{\rm mom}=12$ and $N_{\rm mom}=8$, respectively. The resulting statistics are shown in \cref{tab:statistics}.

For the effective energy fits \cref{eq:eff mass}, we use the two-point function with full statistics. The two-point function with full statistics is calculated using $735$ gauge configurations with $349$ source positions each, for a total of 256,515 measurements.

\section{Renormalization}
The matrix elements of $\mathcal{O}^{\{\mu\nu\rho\}}$ and 
$\mathcal{O}^{\{\mu\nu\rho\lambda\}}$ are renormalized nonperturbatively 
in the RI$'$-MOM scheme~\cite{Martinelli:1994ty} and converted 
perturbatively to the $\overline{\rm MS}$ scheme 
at the scale $\bar{\mu} = 2$ GeV. The renormalization factors for the two- and three-derivative operators with all Lorentz indices different were computed by some of us in Ref.~\cite{Alexandrou:2026nsl} in the context of higher Mellin moments of pion and kaon PDFs. In the present work, we extend this calculation to operators of the form $\mathcal{O}^{\{\mu\mu\rho\lambda\}}$ with $\mu \neq \rho \neq \lambda \neq \mu$ and repeated indices are not summed over. An alternative, traceless basis is provided by $\mathcal{O}^{\{\mu\mu\rho\lambda\}} - \mathcal{O}^{\{\nu\nu\rho\lambda\}}$ with $(\mu,\nu,\rho,\lambda)$ all different and repeated indices again not summed over. Both operator bases support the same irreducible representation of $H(4)$ and therefore have identical renormalization factors. As discussed in Sec.~\ref{sec:Mixing}, this choice of three-derivative operators yields an improved signal-to-noise ratio compared to operators with all Lorentz indices different. It also provides greater flexibility in the extraction of nucleon GFFs, as multiple independent data sets can be constructed through different index assignments. A disadvantage of this choice of operators is the occurrence of mixing with other lattice operators that have the same transformation properties under $H(4)$~\cite{Gockeler:1996mu}. Nevertheless, the mixing pattern does not involve lower-dimensional operators, thus avoiding the need for potentially delicate subtractions associated with power-divergent mixing. 

According to the $H(4)$ classification, $\mathcal{O}_1^{\mu\nu\rho\lambda} \equiv \mathcal{O}^{\{\mu\mu\rho\lambda\}} - \mathcal{O}^{\{\nu\nu\rho\lambda\}}$ mixes with the five operators listed below~\cite{Gockeler:1996mu}:
\begin{eqnarray}
      \mathcal{O}_2^{\mu\nu\rho\lambda} &\equiv& \mathcal{O}_{A}^{\{\rho \mu\}[\rho \nu]} +\mathcal{O}_{A}^{\{\rho \nu\}[\rho \mu]} - \mathcal{O}_{A}^{\{\lambda \mu\}[\lambda \nu]} - \nonumber \\
      && \mathcal{O}_{A}^{\{\lambda \nu\}[\lambda \mu]}, \qquad \\
      \mathcal{O}_3^{\mu\nu\rho\lambda} &\equiv& P_1\Big(\mathcal{O}^{\{\mu \rho\}[\mu \lambda]} + \mathcal{O}^{\{\mu \lambda\}[\mu \rho]} - \mathcal{O}^{\{\nu \rho\}[\nu \lambda]} - \nonumber \\
      && \qquad \mathcal{O}^{\{\nu \lambda\}[\nu \rho]}\Big), \\
      \mathcal{O}_4^{\mu\nu\rho\lambda} &\equiv& P_2\Big(\mathcal{O}^{\{\mu \rho\}[\mu \lambda]} + \mathcal{O}^{\{\mu \lambda\}[\mu \rho]} - \mathcal{O}^{\{\nu \rho\}[\nu \lambda]} - \nonumber  \\
      && \qquad \mathcal{O}^{\{\nu \lambda\}[\nu \rho]}\Big), \\
      \mathcal{O}_5^{\mu\nu\rho\lambda} &\equiv& \mathcal{O}_A^{[\rho \mu][\rho \nu]} +\mathcal{O}_A^{[\rho \nu][\rho \mu]} - \mathcal{O}_A^{[\lambda \mu][\lambda \nu]} - \nonumber \\
      && \mathcal{O}_A^{[\lambda \nu][\lambda \mu]}, \\
      \mathcal{O}_6^{\mu\nu\rho\lambda} &\equiv& P_1\Big(\mathcal{O}^{[\mu \rho][\mu \lambda]} + \mathcal{O}^{[\mu \lambda][\mu \rho]} - \mathcal{O}^{[\nu \rho][\nu \lambda]} - \nonumber \\
      && \qquad \mathcal{O}^{[\nu \lambda][\nu \rho]}\Big),
    \end{eqnarray}
where,
\begin{align}
     \mathcal{O}_A^{\mu \nu \rho \lambda} \equiv \overline{\psi} \, \gamma^{\mu} \gamma^5 \, i \overleftrightarrow{D}{}^{\nu} i \overleftrightarrow{D}{}^{\rho} i\overleftrightarrow{D}{}^{\lambda}\,\psi\,
\end{align}
denotes the corresponding axial-vector three-derivative operator. The mixed-symmetry combinations are defined as:
\begin{eqnarray}
    \mathcal{O}^{\{\mu \rho\}[\nu \lambda]} &\equiv& \frac{1}{2} \Big(\mathcal{O}^{\mu \nu \rho \lambda} + \mathcal{O}^{\rho \nu \mu \lambda} - \mathcal{O}^{\mu \lambda \rho \nu} - \mathcal{O}^{\rho \lambda \mu \nu} \Big), \qquad \\
    \mathcal{O}^{[\mu \rho][\nu \lambda]} &\equiv& \frac{1}{2} \Big(\mathcal{O}^{\mu \nu \rho \lambda} - \mathcal{O}^{\rho \nu \mu \lambda} - \mathcal{O}^{\mu \lambda \rho \nu} + \mathcal{O}^{\rho \lambda \mu \nu} \Big),
\end{eqnarray}
with analogous definitions for the axial-vector operators. $P_1$ and $P_2$ generate specific linear combinations of permutations of the indices $(\mu,\nu,\rho,\lambda)$:
\begin{eqnarray}
    P_1\Big(\mathcal{O}^{\{\mu \rho\}[\nu \lambda]}\Big) &\equiv& \mathcal{O}^{\{\mu \rho\}[\nu \lambda]} -2 (\mathcal{O}^{\nu \mu \rho \lambda} + \mathcal{O}^{\nu \rho \mu \lambda} - \nonumber \\
    && \qquad \qquad \qquad \, \mathcal{O}^{\lambda \mu \rho \nu} - \mathcal{O}^{\lambda \rho \mu \nu}), \\
    P_1\Big(\mathcal{O}^{[\mu \rho][\nu \lambda]}\Big) &\equiv& \mathcal{O}^{[\mu \rho][\nu \lambda]} -2 (\mathcal{O}^{\nu \mu \rho \lambda} - \mathcal{O}^{\nu \rho \mu \lambda} - \nonumber \\
    && \qquad \qquad \qquad \mathcal{O}^{\lambda \mu \rho \nu} + \mathcal{O}^{\lambda \rho \mu \nu}), \\
    P_2\Big(\mathcal{O}^{\{\mu \rho\}[\nu \lambda]}\Big) &\equiv& \mathcal{O}^{\{\mu \rho\}[\nu \lambda]} + \mathcal{O}^{\nu \mu \rho \lambda} + \mathcal{O}^{\nu \rho \mu \lambda} - \nonumber \\
    && \mathcal{O}^{\lambda \mu \rho \nu} - \mathcal{O}^{\lambda \rho \mu \nu} - 3 (\mathcal{O}^{\mu \rho \nu \lambda} + \nonumber \\
    && \mathcal{O}^{\rho \mu \nu \lambda} - \mathcal{O}^{\mu \rho \lambda \nu} - \mathcal{O}^{\rho \mu \lambda \nu}).
\end{eqnarray}
Mixing between vector and axial-vector three-derivative operators is not protected by parity for the specific index combinations considered above. This follows from the fact that the relevant vector and axial-vector operators contain opposite pairings of unrepeated indices, with only one index among ($\mu,\nu,\rho,\lambda$) being temporal.
The renormalized operator $\mathcal{O}_1^{\mu\nu\rho\lambda,R}$ is then obtained as a linear combination of the six bare operators:
\begin{equation}
    \mathcal{O}_1^{\mu\nu\rho\lambda,R} = Z_{11} \mathcal{O}_1^{\mu\nu\rho\lambda} + \sum_{i=2}^6 Z_{1i} \mathcal{O}^{\mu\nu\rho\lambda,R}_i.
\end{equation}
A complete treatment of renormalization requires the determination of the matrix elements of all six operators. In this analysis, however, we neglect the mixing contributions and approximate the renormalization as multiplicative,
\begin{equation}
\mathcal{O}_1^{\mu\nu\rho\lambda,R} \approx Z_{11} \,\mathcal{O}_1^{\mu\nu\rho\lambda}.
\end{equation}
The associated systematic uncertainty from the omitted term $\sum_{i=2}^{6} Z_{1i} \, \mathcal{O}_i^{\mu\nu\rho\lambda}$ can be estimated by comparing the form factor $A_{40}$ extracted from the renormalized matrix elements of $\mathcal{O}^{\{\mu\nu\rho\lambda\}}$ and of $\mathcal{O}_1^{\mu\nu\rho\lambda}$. The two determinations of $A_{40}$ are expected to agree up to $\mathcal{O}(a^2)$ discretization effects in the twisted-mass formulation. From our previous studies, $\mathcal{O}(a^2)$ effects are expected to be small at the present lattice spacing and below the level of our current statistical uncertainties~\cite{Alexandrou:2026soz}. 
We will thus neglect this mixing and  determine the renormalization factor $Z_{11} \equiv Z_{{\rm VDDD}_2}$ nonperturbatively following the procedure described in, e.g., Refs.~\cite{ExtendedTwistedMass:2024kjf,Alexandrou:2024ozj,Alexandrou:2025vto}.

It should be noted that, for finite $Q^2$, the matrix elements of $\mathcal{O}^{\{\mu\nu\rho\}}$ and 
$\mathcal{O}^{\{\mu\nu\rho\lambda\}}$ exhibit additional mixing with matrix elements of total-derivative operators~\cite{Gracey:2009da}, both in the continuum and on the lattice. In this case, the RI$'$/MOM scheme does not provide access to the offdiagonal mixing coefficients of the renormalization matrix. Determining these coefficients requires alternative schemes such as RI$'$/SMOM~\cite{Gracey:2011zg,Kniehl:2020nhw}, where the renormalization conditions are imposed on vertex functions with nonvanishing momentum transfer $q$. The offdiagonal mixing coefficient for the two-derivative operator has been calculated at one-loop order in lattice perturbation theory in Ref.~\cite{Gockeler:2004xb} and found to be very small compared to the diagonals. Thus, in this first  analysis of such operators, we ignore the offdiagonal mixing contributions and consider multiplicative renormalization. 

The RI$'$/MOM scheme is defined on amputated vertex functions of the operator under study with external offshell quark states in Landau gauge:
\begin{equation}
    \Lambda (p) = \frac{a^{12}}{V} \sum_{x,y,z} e^{-i p (x-y)} \langle q (x) \mathcal{O}_1^{\mu\nu\rho\lambda} (z) \bar{q} (y) \rangle_{\rm amp.}, 
\end{equation}
In the continuum limit, $\Lambda (p)$ is decomposed into two independent structures allowed by rotational symmetry~\cite{Gracey:2006zr}:
\begin{eqnarray}
      \Lambda (p) &=& - i \left(\gamma^{\{\mu} p^{\mu} p^{\rho} p^{\lambda \}} - \gamma^{\{\nu} p^{\nu} p^{\rho} p^{\lambda \}}\right) \ \Sigma_1 (p^2) + \nonumber \\
      && \frac{\slashed{p}}{p^2} \left((p^\mu)^2 -  (p^\nu)^2 \right) p^\rho p^\lambda \ \Sigma_2 (p^2),
    \end{eqnarray}
    where $\Sigma_1 (p^2) = 1 + \mathcal{O} (\alpha_s)$, and $\Sigma_2 (p^2) = \mathcal{O} (\alpha_s)$. In continuum regularizations, the renormalization conditions are typically defined in terms of the first form factor, $\Sigma_1(p^2)$~\cite{Gracey:2006zr}, which can be isolated by applying a suitable projector to the vertex functions. On the lattice, however, the projector must also be orthogonal to the tree-level structures of $\mathcal{O}_2 - \mathcal{O}_6$ to ensure that mixing effects are eliminated, at least up to $\mathcal{O}(\alpha_s)$. To this end, we impose the following condition to extract $Z_{{\rm VDDD}_2}$:
    \begin{equation}
      {(Z_q^{{\rm RI}'})}^{-1} Z_{{\rm VDDD}_2}^{{\rm RI}'} \, \frac{1}{12 N_p} \sum_{\mu<\nu} \sum_{\rho<\lambda} {\rm Tr} \left[ \Lambda (p) \, P^{\mu\nu\rho\lambda} \right]\Big|_{p^2 = \mu_0^2} = 1,
    \end{equation}
where 
\begin{equation}
P^{\mu\nu\rho\lambda} = \frac{i}{\tilde{p}^\rho \tilde{p}^\lambda} \frac{1}{(\tilde{p}^\mu)^2 - (\tilde{p}^\nu)^2} \Big[\slashed{\tilde{p}} - \frac{\tilde{p}^2}{2} \Big(\frac{\gamma^\mu}{\tilde{p}^\mu} + \frac{\gamma^\nu}{\tilde{p}^\nu}\Big)\Big],
\end{equation}
and $\tilde{p}^\mu \equiv \sin(a p^\mu)$, $\slashed{\tilde{p}} \equiv \sum_\mu \gamma^\mu \tilde{p}^\mu$. ($\tilde{p}^\mu$, $\tilde{p}^\nu$, $\tilde{p}^\rho$, $\tilde{p}^\lambda$) are all strictly nonzero. The sums over $(\mu,\nu,\rho,\lambda)$ run only over the $N_p$ terms satisfying $(\tilde{p}^\mu)^2 \neq (\tilde{p}^\nu)^2$ and all indices are distinct. $\mu_0$ represents the RI$'$/MOM scale. $Z_q^{{\rm RI}'}$ is the renormalization factor of the quark field defined by~\cite{ExtendedTwistedMass:2021gbo}:
\begin{equation}
     Z_q^{{\rm RI}'} = \frac{1}{12} \sum_{\mu} {\rm Tr} \left[S^{-1} (p) \cdot \frac{-i \ \gamma^\mu}{4 {\tilde{p}}^\mu}\right] \Big|_{p^2 = \mu_0^2},
\end{equation}
where $S(p)$ is the quark propagator in the momentum space.  
 \begin{figure}[h!]
\centering
\includegraphics[width=\columnwidth]{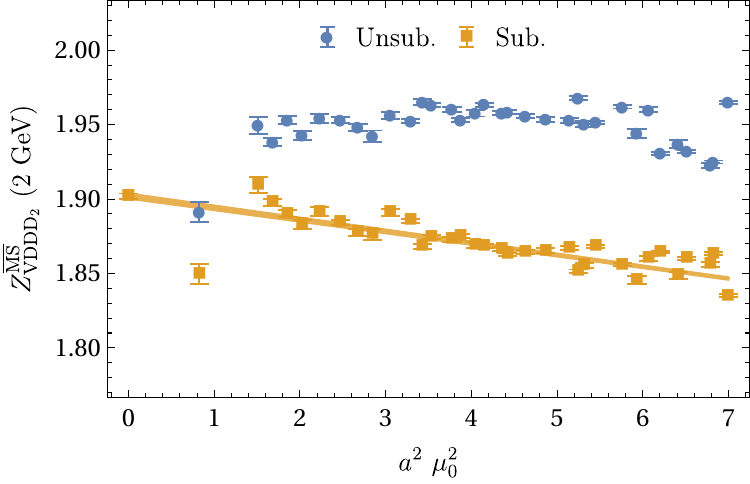} 
    \caption{$Z_{{\rm VDDD}_2}^{\overline{\rm MS}}$ as a function of $a^2 \mu_0^2$ at the reference scale of 2 GeV. The data are given with (Sub.) and without (Unsub.) subtracting one-loop artifacts. A linear fit $c_0 + c_1 a^2 \mu_0^2$ is employed in the subtracted data and the extrapolated values $c_0$ are given at $a^2 \mu_0^2 = 0$.}
    \label{fig:Zfactors}
\end{figure}
The vertex functions $\Lambda (p)$ and quark propagators $S(p)$ are calculated using Landau gauge fixed momentum sources~\cite{Gockeler:1998ye}, which leads to high statistical accuracy using only 30 configurations. Since we employ a mass-independent renormalization scheme, we simulate four $N_f=4$ ensembles~\cite{Alexandrou:2024ozj} with mass-degenerate quarks  at the same $\beta$ value as the  ensemble used in our analysis of the matrix elements. Each of the four ensembles is simulated at a different value of the twisted-mass parameter $\mu_{\rm sea}$, or equivalently ``pion'' mass, and are used in order to take the chiral limit. 
The dependence on $\mu_{\rm sea}$ is found to be mild, consistent with our previous studies of similar operators~\cite{Alexandrou:2021mmi,Alexandrou:2026nsl}. We remove this dependence by performing a linear fit in $\mu_{\rm sea}$. 

To minimize rotational $O(4)$ breaking lattice effects, we consider momenta close to the body-diagonal direction by imposing $\sum_{\mu} (p^\mu)^4/[\sum_\mu (p^\mu)^2]^2 < 0.3$. Additionally, we improve our nonperturbative values by subtracting one-loop lattice artifacts from both $Z_q$ and $\Lambda (p)$. The artifacts are computed in lattice perturbation theory to all orders in the lattice spacing by extending our improvement program~\cite{Alexandrou:2015sea} to the specific three-derivative operators. 

After chiral extrapolation, the renormalization factor is converted to the $\overline{\rm MS}$ scheme and evolved at the reference scale 2 GeV, using an intermediate Renormalization Group Invariant (RGI) scheme~\cite{Gockeler:2010yr}. The anomalous dimension of the operator $\mathcal{O}_1^{\mu\nu\rho\lambda}$, which enters the conversion and evolution, can be derived to four loops in perturbation theory by using the results of Refs.~\cite{Gracey:2006zr,Baikov:2015tea,Herzog:2018kwj}.

\begin{figure}[ht!]
\centering
\includegraphics[width=\columnwidth]{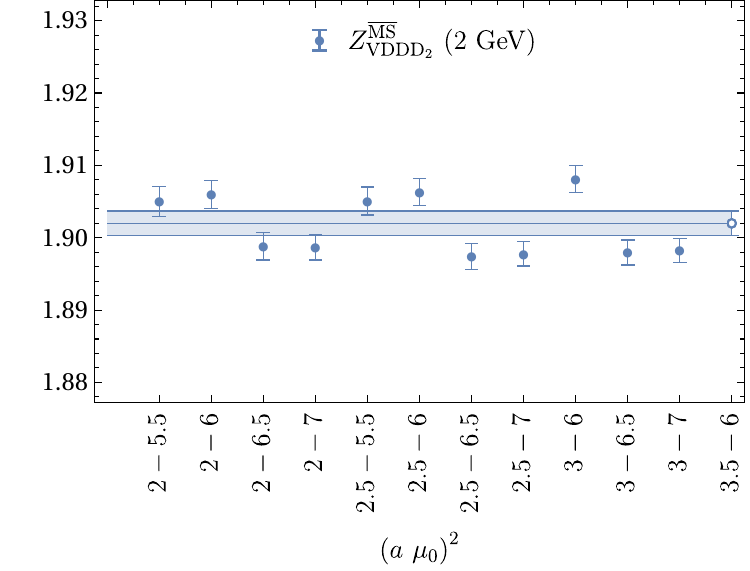} 
    \caption{Extrapolated values of $Z_{{\rm VDDD}_2}^{\overline{\rm MS}}$ at $a^2 \mu_0^2 = 0$ from momentum fits across multiple ranges together with the AIC-averaged value (band).}
    \label{fig:AIC}
\end{figure} 

We apply a linear fit in $a^2 \mu_0^2$ on the renormalization factors in the $\overline{\rm MS}$ to eliminate any residual dependence on the RI$'$/MOM scale resulting from discretization effects. 
We employ several fit ranges within $2 \leq (a\mu_0)^2 \leq 7$. The extrapolated values at $\mu_0 = 0$ from all fits are combined using model averaging based on the Akaike Information Criterion (AIC)~\cite{Jay:2020jkz}. Momenta with $(a\mu_0)^2 < 2$ are excluded from the analysis, as they may suffer from significant hadronic contamination, as well as the perturbative conversion is not reliable in this low-momentum region. 
Fig.~\ref{fig:Zfactors} displays  the linear fit to the results on  $Z_{{\rm VDDD}_2}$ that corresponds to the  fit range with the highest AIC weight. In the figure, we also include results without subtracting one-loop cut-off artifacts shown in order to illustrate the benefit of our subtraction method. 
In Fig.~\ref{fig:AIC}, we show the results from all fits together with the AIC-averaged value.

The final value for $Z_{{\rm VDDD}_2}$ is given below. For completeness, we also provide the renormalization factors of $\mathcal{O}^{\{\mu\nu\rho\}}$ and 
$\mathcal{O}^{\{\mu\nu\rho\lambda\}}$, denoted as $Z_{\rm VDD}$ and $Z_{\rm VDDD}$, respectively, taken from Ref.~\cite{Alexandrou:2026nsl}:  
\begin{eqnarray}
    Z_{\rm VDD}^{\overline{\rm MS}} \, (2 \, {\rm GeV}) &=& 1.4639(11)(03), \\
    Z_{\rm VDDD}^{\overline{\rm MS}} \, (2 \, {\rm GeV}) &=& 1.8831(19)(10), \\
    Z_{{\rm VDDD}_2}^{\overline{\rm MS}} \, (2 \, {\rm GeV}) &=& 1.9020(17)(34).
\end{eqnarray}
The number in the first (second) parenthesis corresponds to the statistical (systematic) uncertainty. The systematic uncertainty is determined from the AIC procedure.

\section{Results}
\subsection{Forward limit}
We first present our analysis  for the Mellin moments $\langle x^2\rangle$ and $\langle x^3 \rangle$, which do not require a momentum transfer to be evaluated. For the Mellin moments we follow the convention of Ref.~\cite{Lin:2017snn} and denote the third and fourth moment by $\langle x^2 \rangle_{q_-}$ and $\langle x^3 \rangle_{q_+}$, respectively. The sign of $q_\pm$ comes from integrating the anti-quark distribution. Rewriting \cref{eq: momentsGFF} explicitly in quark and anti-quark one finds for the $n^{\rm th}$ Mellin moment
\begin{align}
    \langle x^{n-1} \rangle = \int_0^1 {\rm d}x\, x^{n-1} [q(x)-(-1)^{n-1} \bar{q}(x)] \,.
    \label{eq:defMellinmoment}
\end{align} So for moments with odd $n$ the anti-quark contribution gets added to the quark contribution ($q_+$), while for moments with even $n$ it gets subtracted ($q_-$). For compactness, we will refer to the flavor $q$ when referring to a specific Mellin moment or a GFF and omit the flavor otherwise.

\begin{figure}[h!]
\centering
    \includegraphics[width=\linewidth]{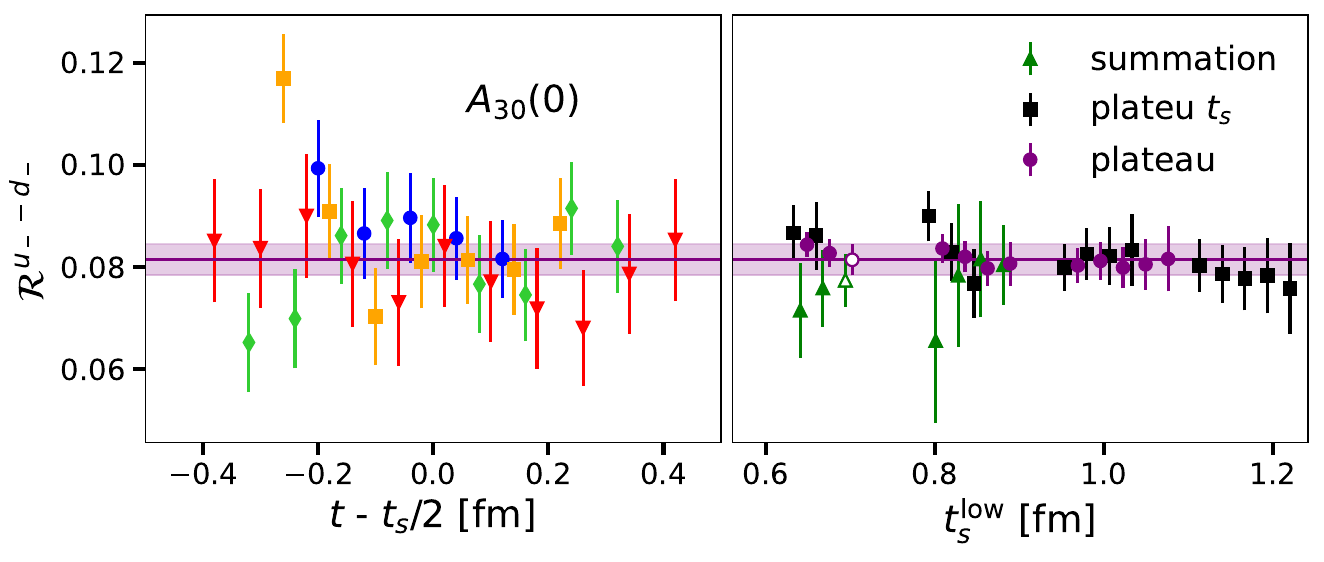}
     \caption{Left: We show results on the ratio of \cref{eq:Ratio} yielding the isovector GFF $A^{u_--d_-}_{30}(0)=\langle x^2\rangle_{u_--d_-}$. Results are  shown for $t_s=8a$ (blue circles), $10a$ (orange squares), $12a$ (green diamonds) and $14a$ (red downwards pointing triangles).  Right: We show results as a function of the smallest time separation  $t_s^{\rm low}$  used in the fits.  The green triangles are results from the summation method, the purple points are from the plateau fit to all separations starting from the given $t_s^{\rm low}$. The black squares show the plateau fit separately for  each time separation $t_s$. Each bundle of points shows the change as  points are eliminated symmetrically near the source and sink starting with $t_{\rm cut}=2 a$. The open purple point gives the selected value and the purple band through both panels is the corresponding error band. We take the difference between the open purple point and the open green triangle as the systematic error.}
    \label{fig:IsovectorTwoDFL}
\end{figure}

\begin{figure}[h!]
\centering
    \includegraphics[width=\linewidth]{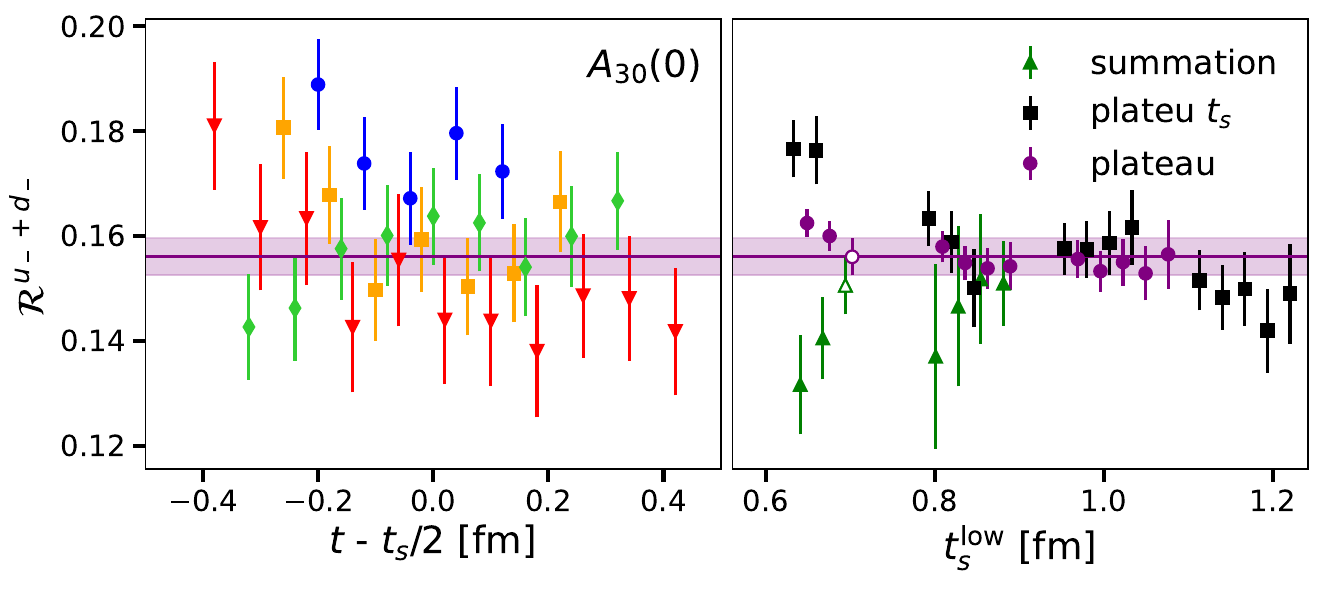}
     \caption{Same as \cref{fig:IsovectorTwoDFL}, but for the isoscalar GFF $A_{30}^{u_-+d_-}(0)=\langle x^2 \rangle_{u_-+d_-}$.}
    \label{fig:IsoscalarTwoDFL}
\end{figure}

In \cref{fig:IsovectorTwoDFL,fig:IsoscalarTwoDFL}, we show  results on the isovector and isoscalar ratio, respectively, for the two-derivative operator case. 
\begin{figure}[h!]
\centering 
    \includegraphics[width=\linewidth]{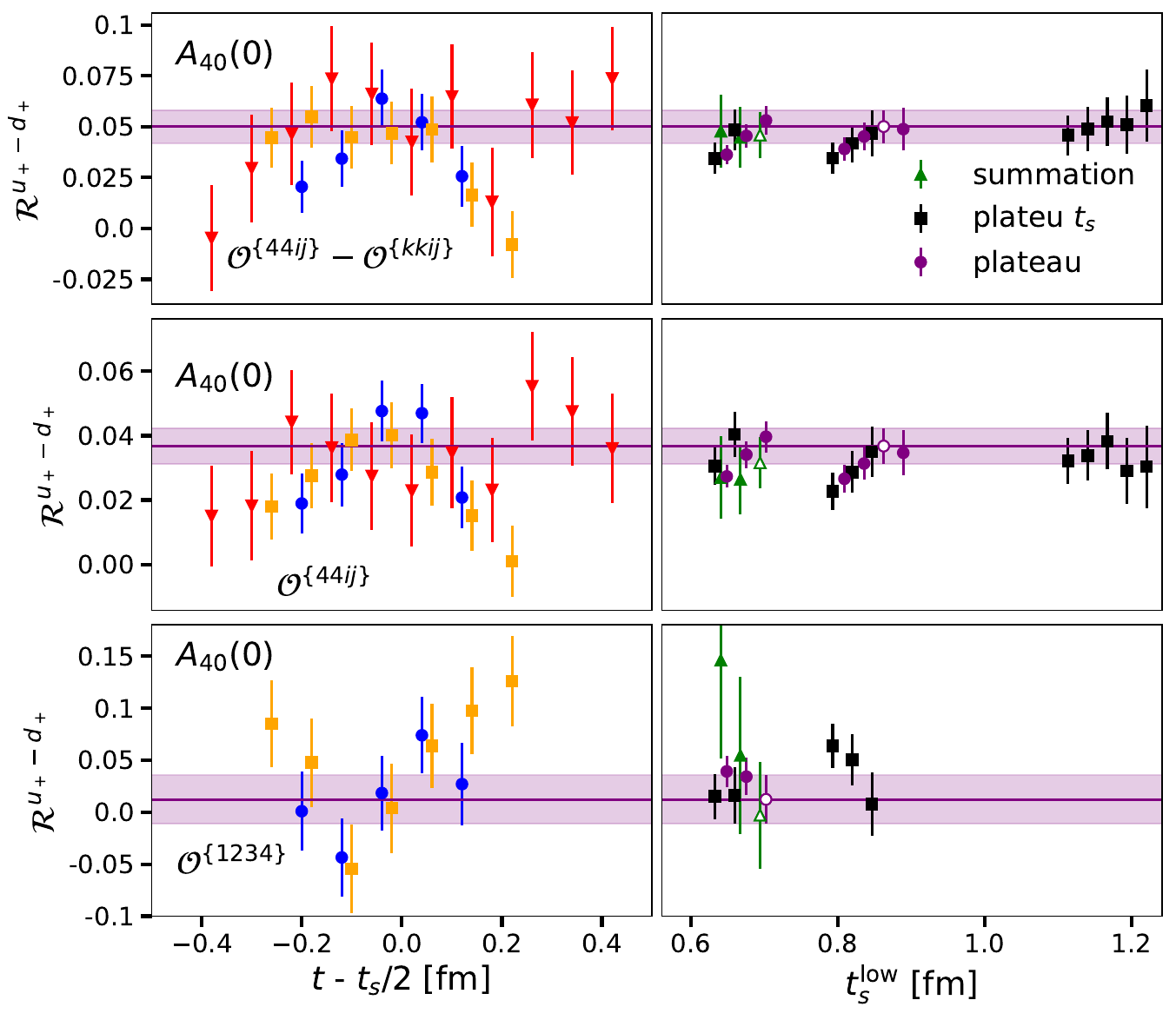}
     \caption{We show results on the ratio of \cref{eq:Ratio} yielding the isovector GFF $A^{u_+-d_+}_{40}(0)=\langle x^3\rangle_{u_+-d_+}$. In the top panels, we show the ratio when using the operator $\mathcal{O}^{\{44ij\}}-\mathcal{O}^{\{kkij\}}$, in the middle panels, the operator $\mathcal{O}^{\{44ij\}}$ and in the bottom panels, $\mathcal{O}^{\{1234\}}$. The rest of the notation is the same as in \cref{fig:IsovectorTwoDFL}. For $\mathcal{O}^{\{1234\}}$ we only computed the two smaller time separations since the errors are large.}
    \label{fig:IsovectorThreeDFL}
\end{figure}
\begin{figure}[h!]
\centering 
    \includegraphics[width=0.95\linewidth]{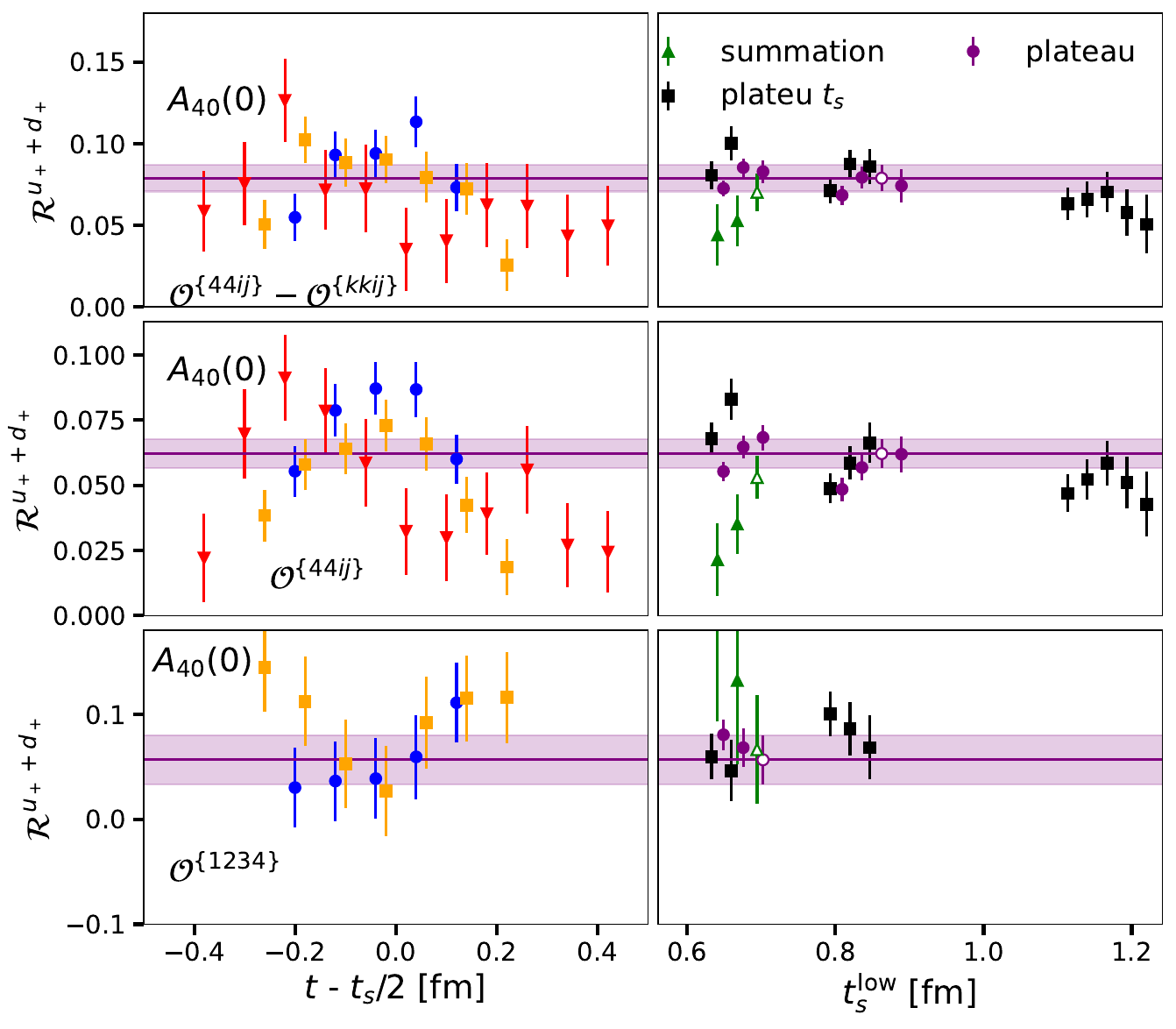}
     \caption{Same as \cref{fig:IsovectorThreeDFL}, but for the isoscalar GFF $A^{u_++d_+}_{40}(0)=\langle x^3\rangle_{u_++d_+}$.}
    \label{fig:IsoscalarThreeDFL}
\end{figure} 
\begin{figure}[htb]
\centering
    \includegraphics[width=0.95\linewidth]{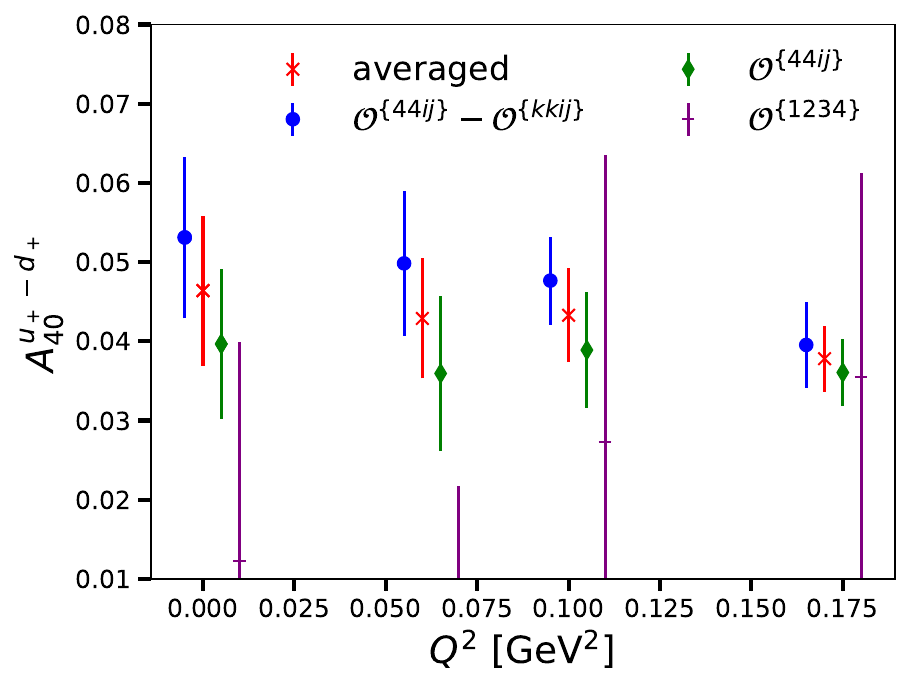}
    \includegraphics[width=0.95\linewidth]{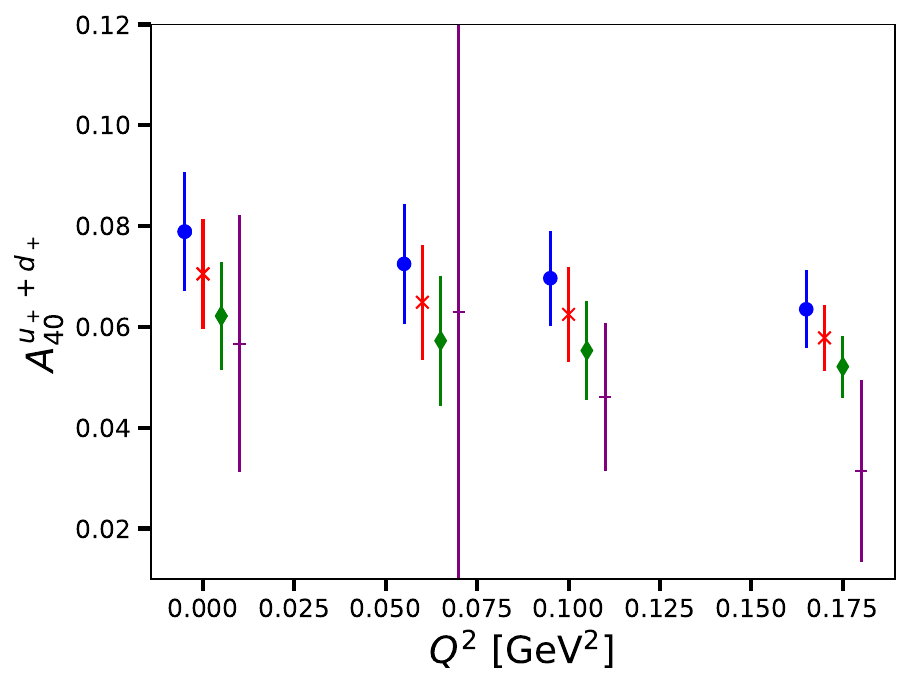}
     \caption{Results on the isovector $A_{40}^{u_+-d_+}(Q^2)$ (top) and isoscalar $A_{40}^{u_++d_+}(Q^2)$ (bottom) as a function of $Q^2$ for the different three-derivative operators. The blue points are from using matrix element of  the operator $\mathcal{O}^{44ij}-\mathcal{O}^{kkij}$, the green diamonds are from $\mathcal{O}^{\{44ij\}}$ and the purple crosses are from $\mathcal{O}^{\{1234\}}$. The red crosses are the average of the operators $\mathcal{O}^{44ij}-\mathcal{O}^{kkij}$ and $\mathcal{O}^{\{44ij\}}$. The different operators are  shifted for better readability.}
    \label{fig:ThreeD_A40_Q2_comp}
\end{figure}
\begin{figure}[h!]
\centering
    \includegraphics[width=\linewidth]{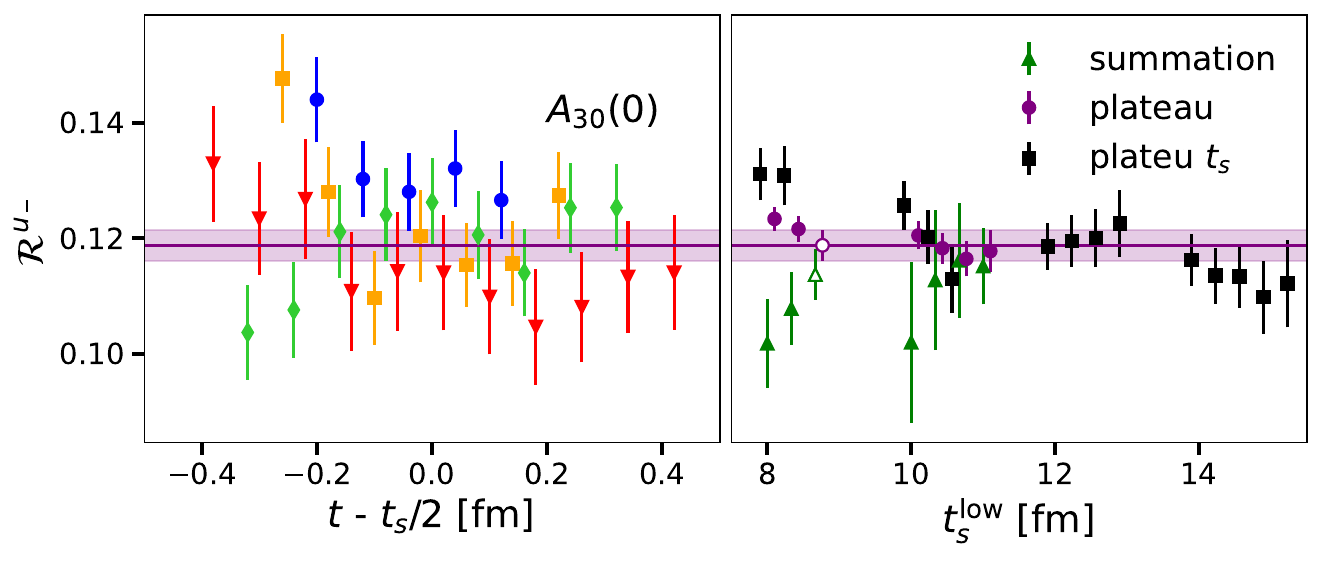}
    \includegraphics[width=\linewidth]{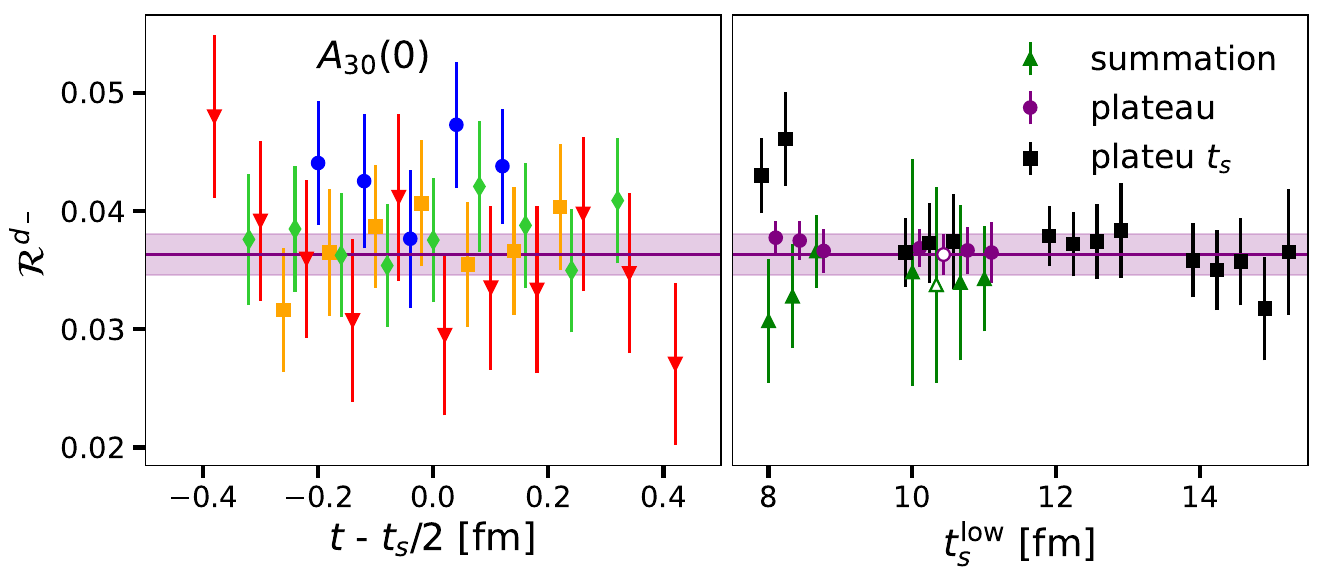}
     \caption{Same as \cref{fig:IsovectorTwoDFL}, but for the up GFF $A_{30}^{u_-}(0)=\langle x^2 \rangle_{u_-}$ (top) and down GFF $A_{30}^{d_-}(0)=\langle x^2 \rangle_{d_-}$ (bottom).}
    \label{fig:UpdnTwoDFL}
\end{figure}

\begin{figure*}[hbt]
    \centering
    \includegraphics[width=0.49\linewidth]{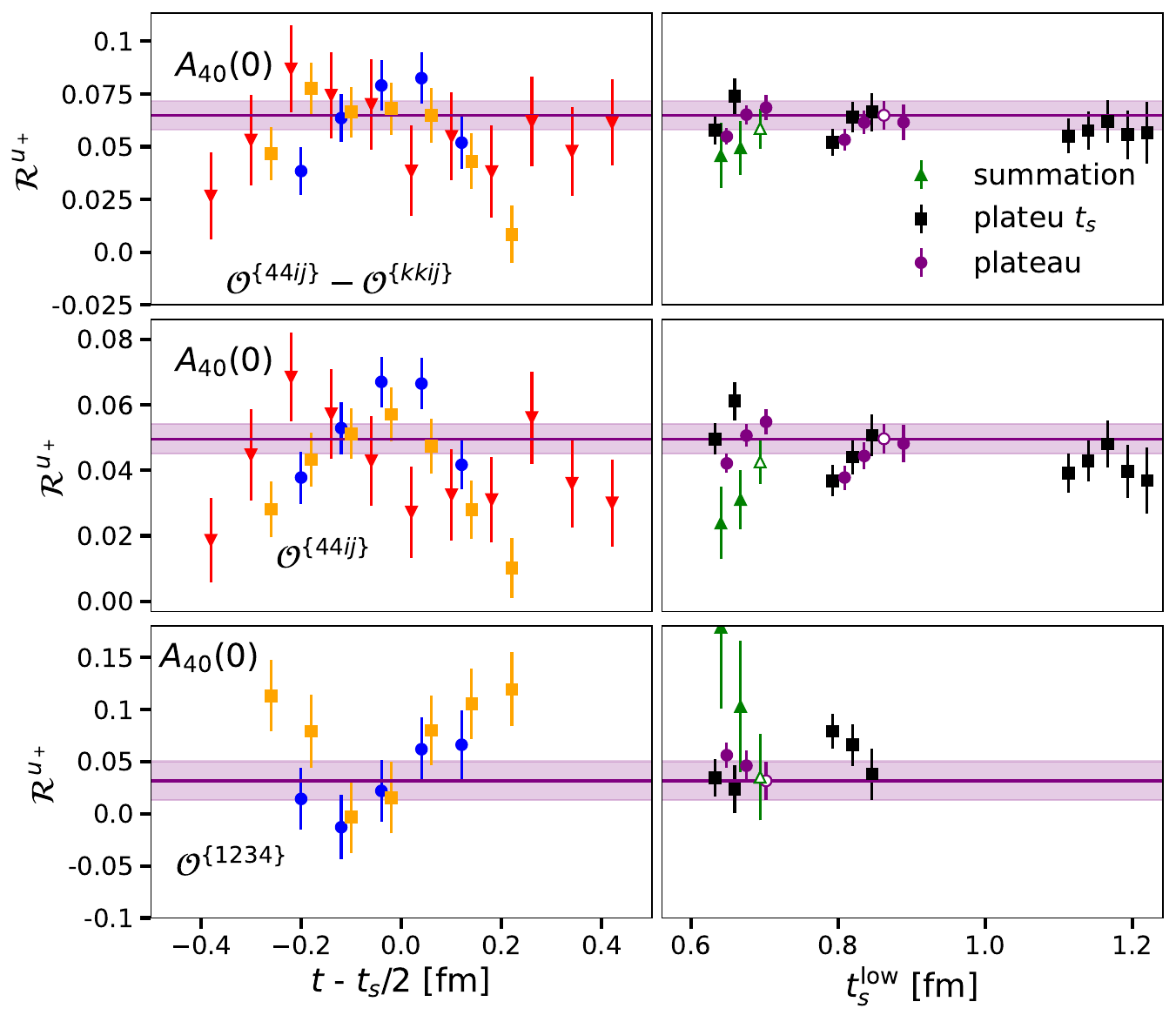}\hfill
    \includegraphics[width=0.49\linewidth]{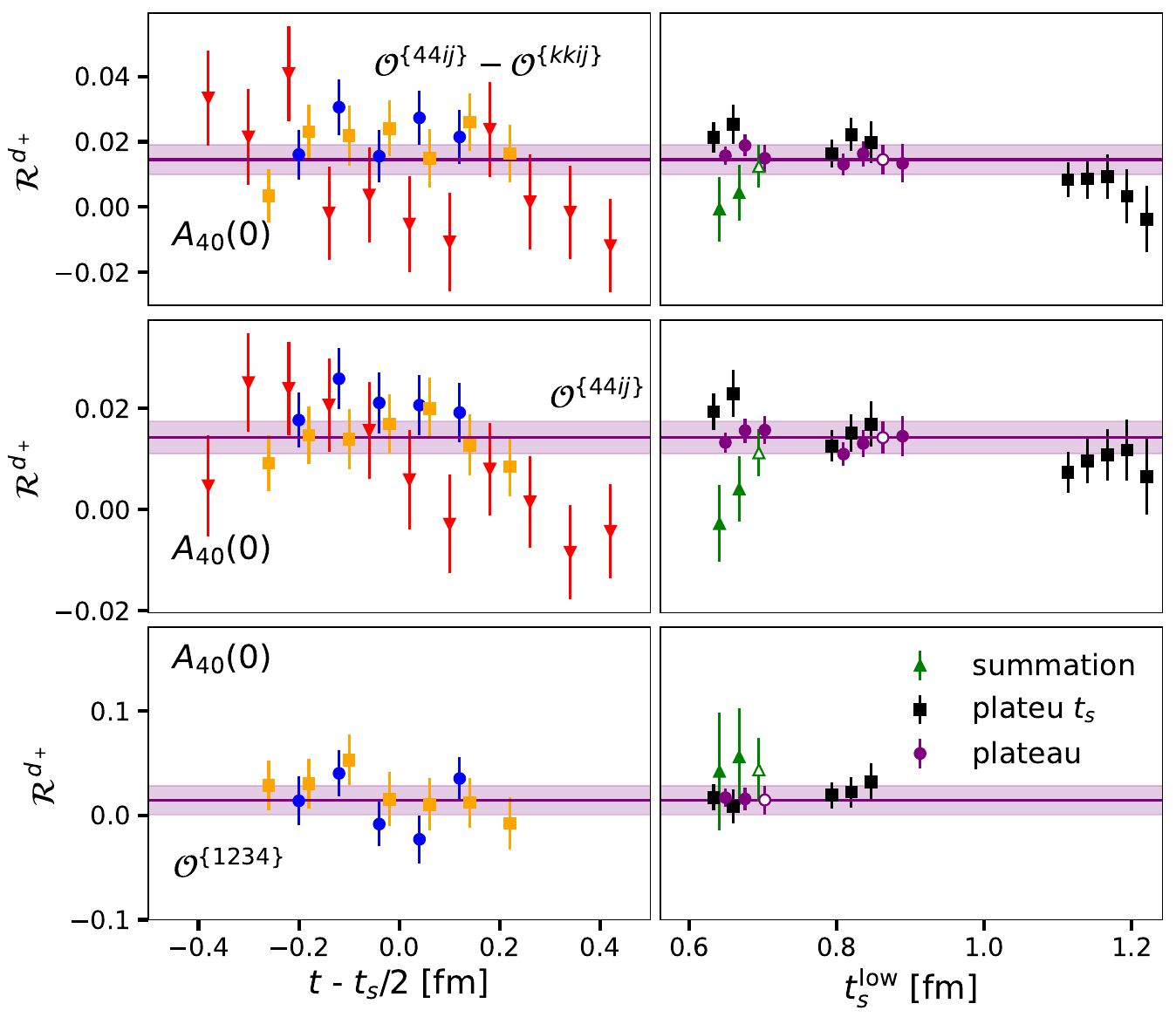}\hfill
    \caption{Same as \cref{fig:IsovectorThreeDFL}, but for the up GFF $A^{u_+}_{40}(0)=\langle x^3\rangle_{u_+}$ (left) and the down GFF $A^{d_+}_{40}(0)=\langle x^3\rangle_{d_+}$ (right).}
    \label{fig:Updn ThreeD GFFs}
\end{figure*}

\begin{figure*}[htb]
    \centering
    \includegraphics[width=0.49\linewidth]{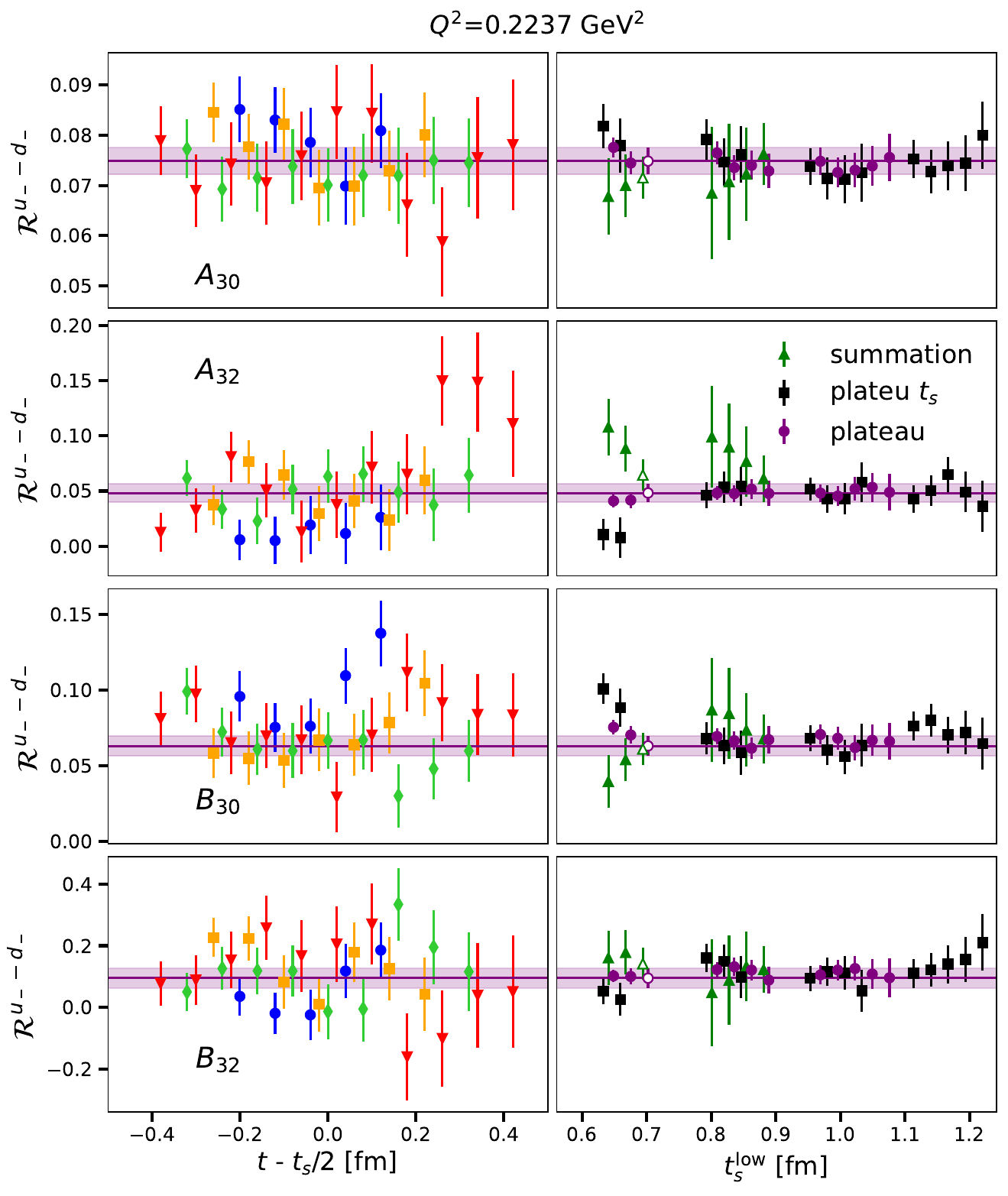}\hfill
    \includegraphics[width=0.49\linewidth]{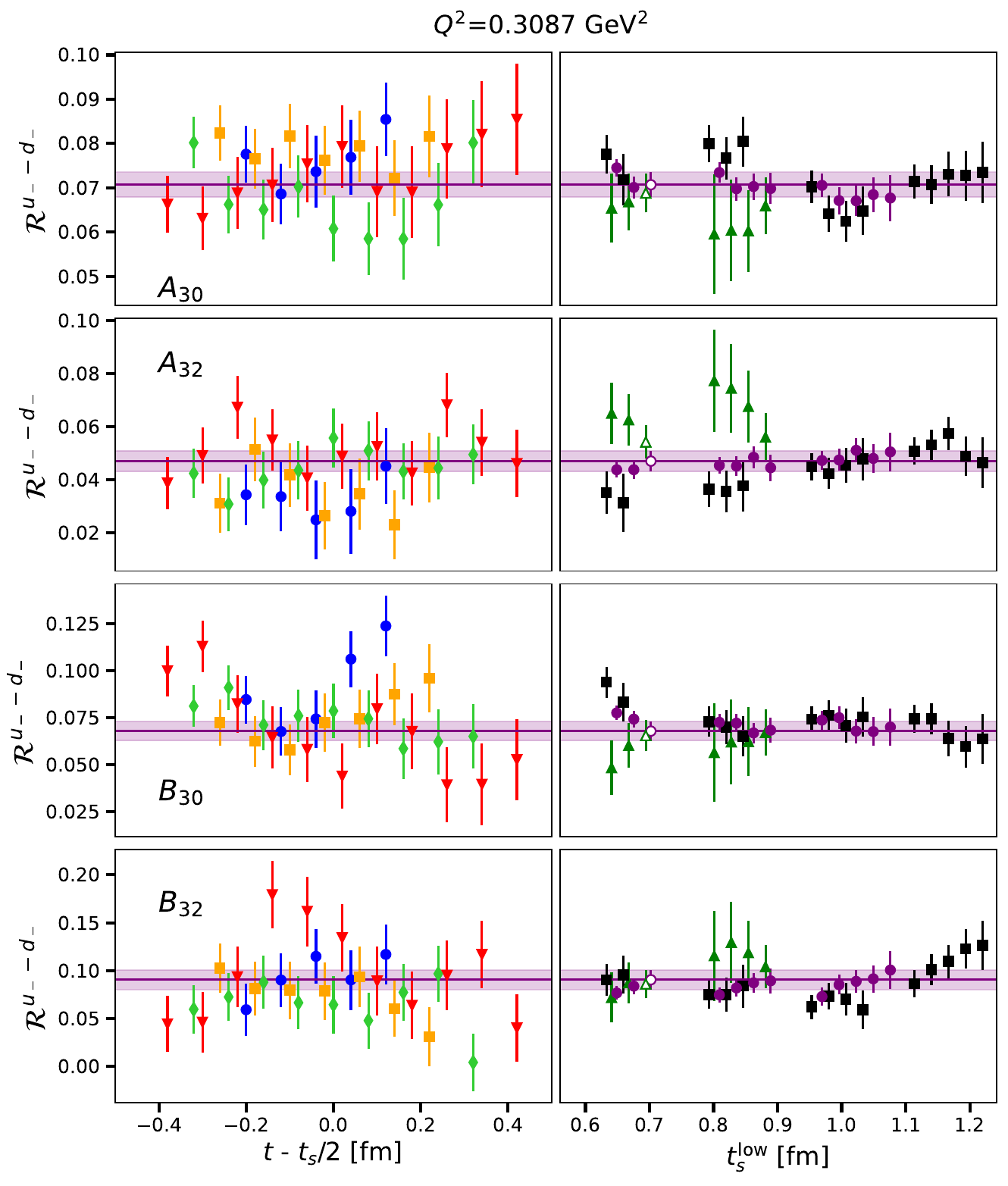}\hfill
    \caption{Results for the four isovector GFFs $A_{30}^{u_--d_-}(Q^2)$, $A_{32}^{u_--d_-}(Q^2)$, $B_{30}^{u_--d_-}(Q^2)$ and $B_{32}^{u_--d_-}(Q^2)$ at $Q^2=-(p^\prime-p)^2=0.2237 \GeV^2$ (left) and $Q^2=0.3087 \GeV^2$. The notation is the same as  in \cref{fig:IsovectorTwoDFL}.}
    \label{fig:TwoD GFFs}
\end{figure*}

\begin{figure*}[hbt]
    \centering
    \includegraphics[width=0.49\linewidth]{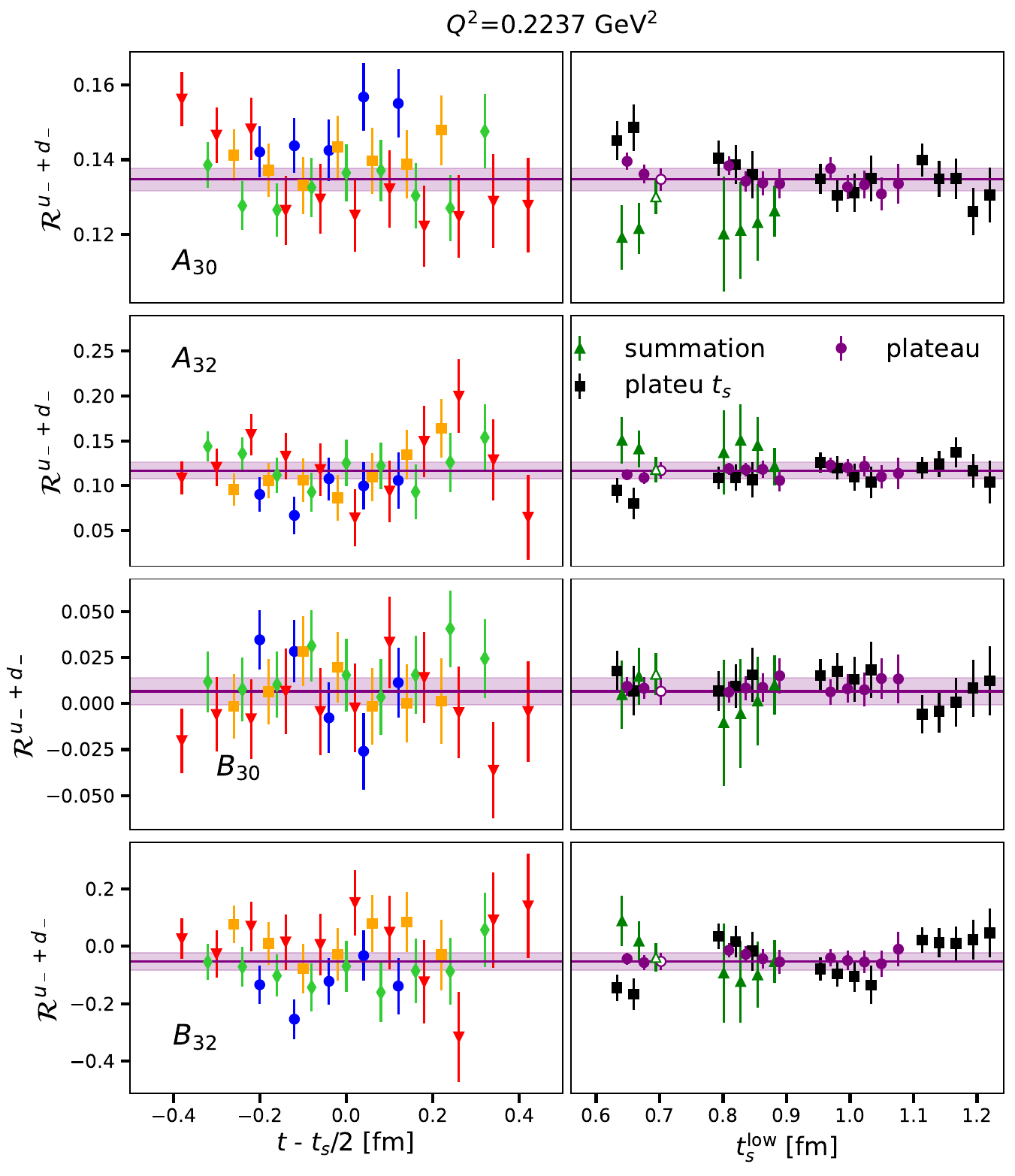}\hfill
    \includegraphics[width=0.49\linewidth]{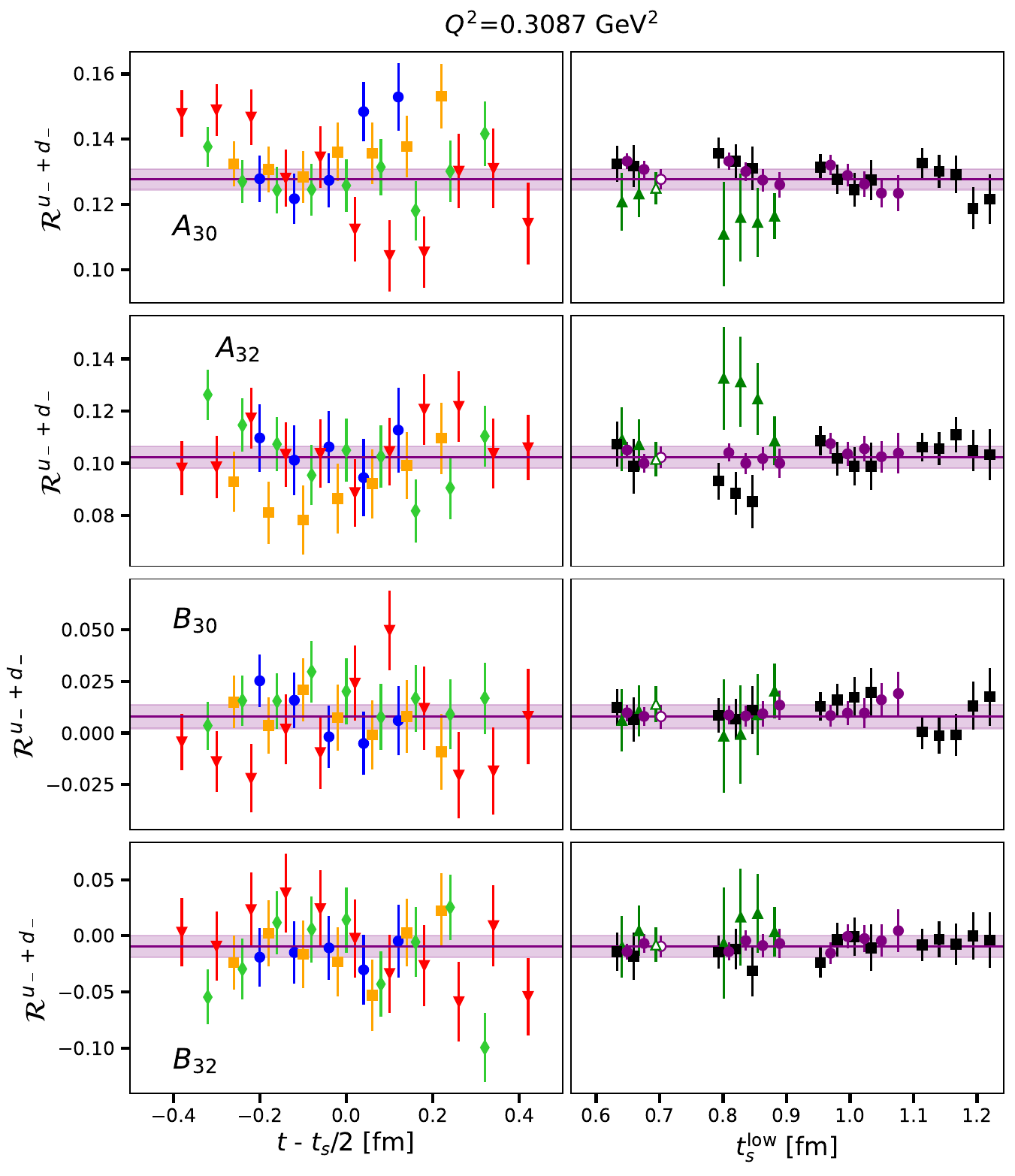}\hfill
    \caption{Same as \cref{fig:TwoD GFFs}, but for the isoscalar GFFs $A_{30}^{u_-+d_-}(Q^2)$, $A_{32}^{u_-+d_-}(Q^2)$, $B_{30}^{u_-+d_-}(Q^2)$ and $B_{32}^{u_-+d_-}(Q^2)$. }
    \label{fig:TwoD GFFs isos}
\end{figure*}
\begin{figure*}[htb]
    \centering
    \includegraphics[width=0.49\linewidth]{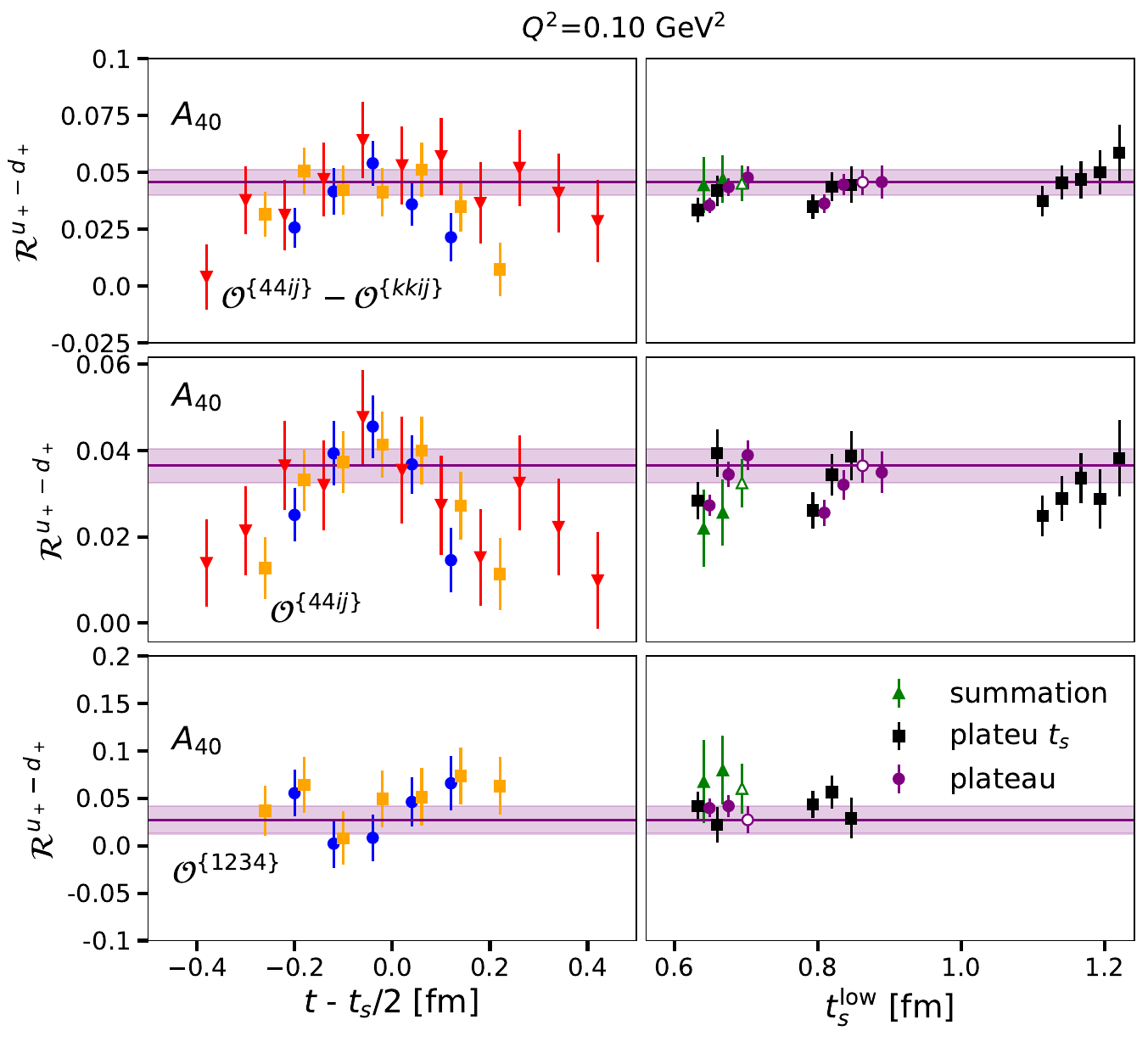}\hfill
    \includegraphics[width=0.49\linewidth]{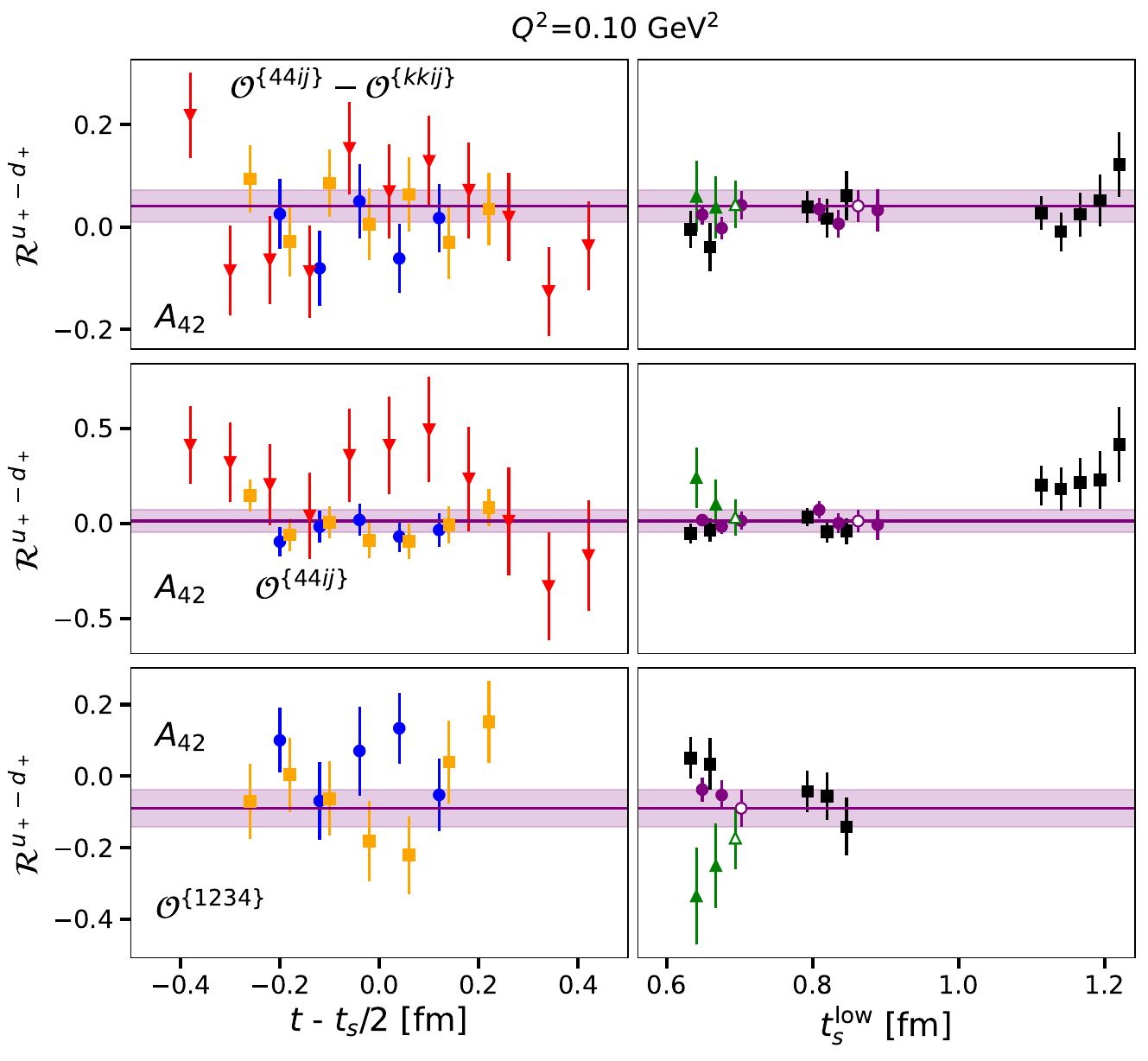}\hfill
    \caption{Three derivative isovector GFFs $A_{40}^{u_+-d_+}(Q^2)$ (left) and $A_{42}^{u_+-d_+}(Q^2)$ (right) at $Q^2=0.1 \GeV^2$.}
    \label{fig:ThreeD GFFs}
\end{figure*}
\begin{figure*}[htb]
    \centering
    \includegraphics[width=0.49\linewidth]{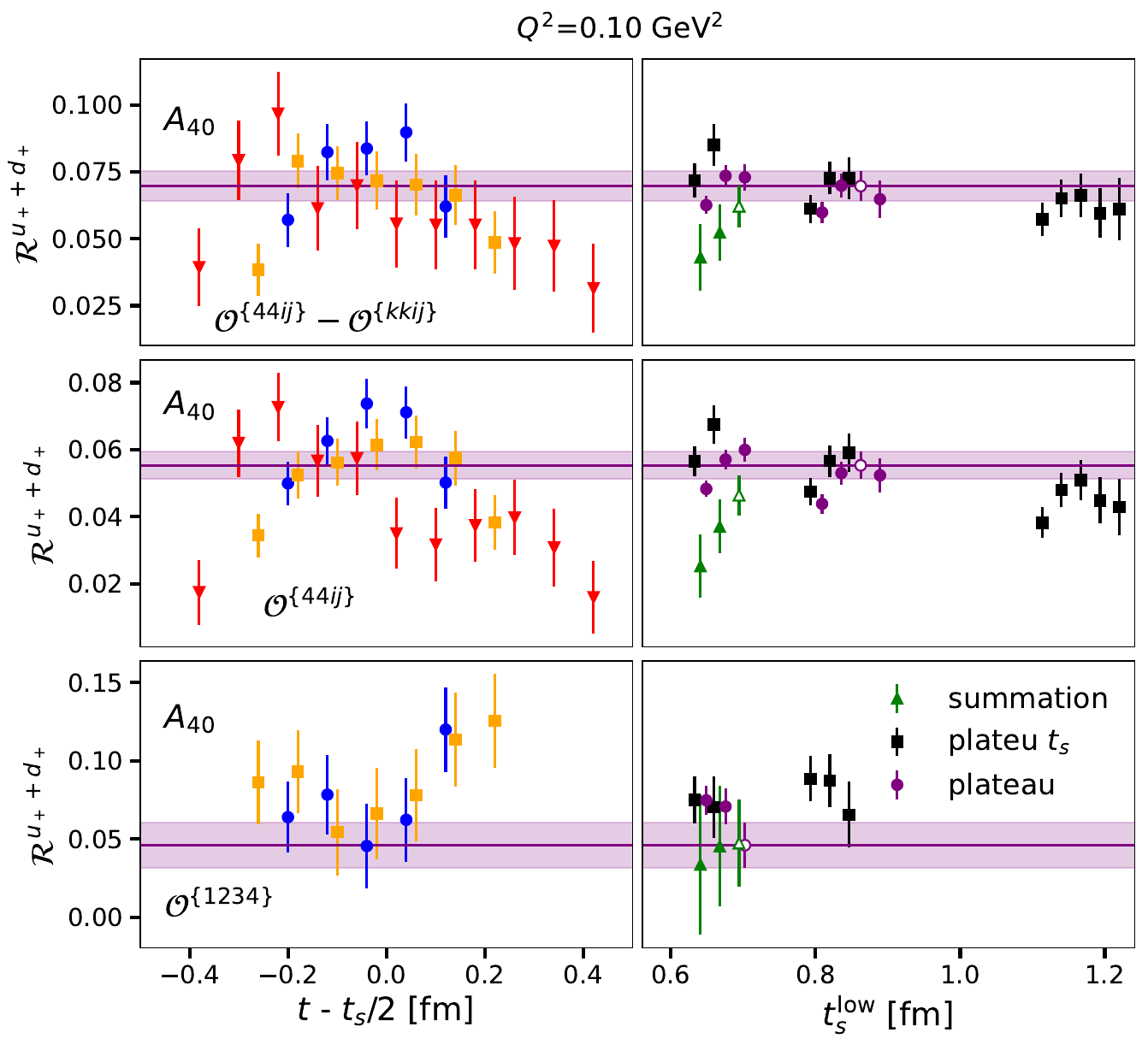}\hfill
    \includegraphics[width=0.49\linewidth]{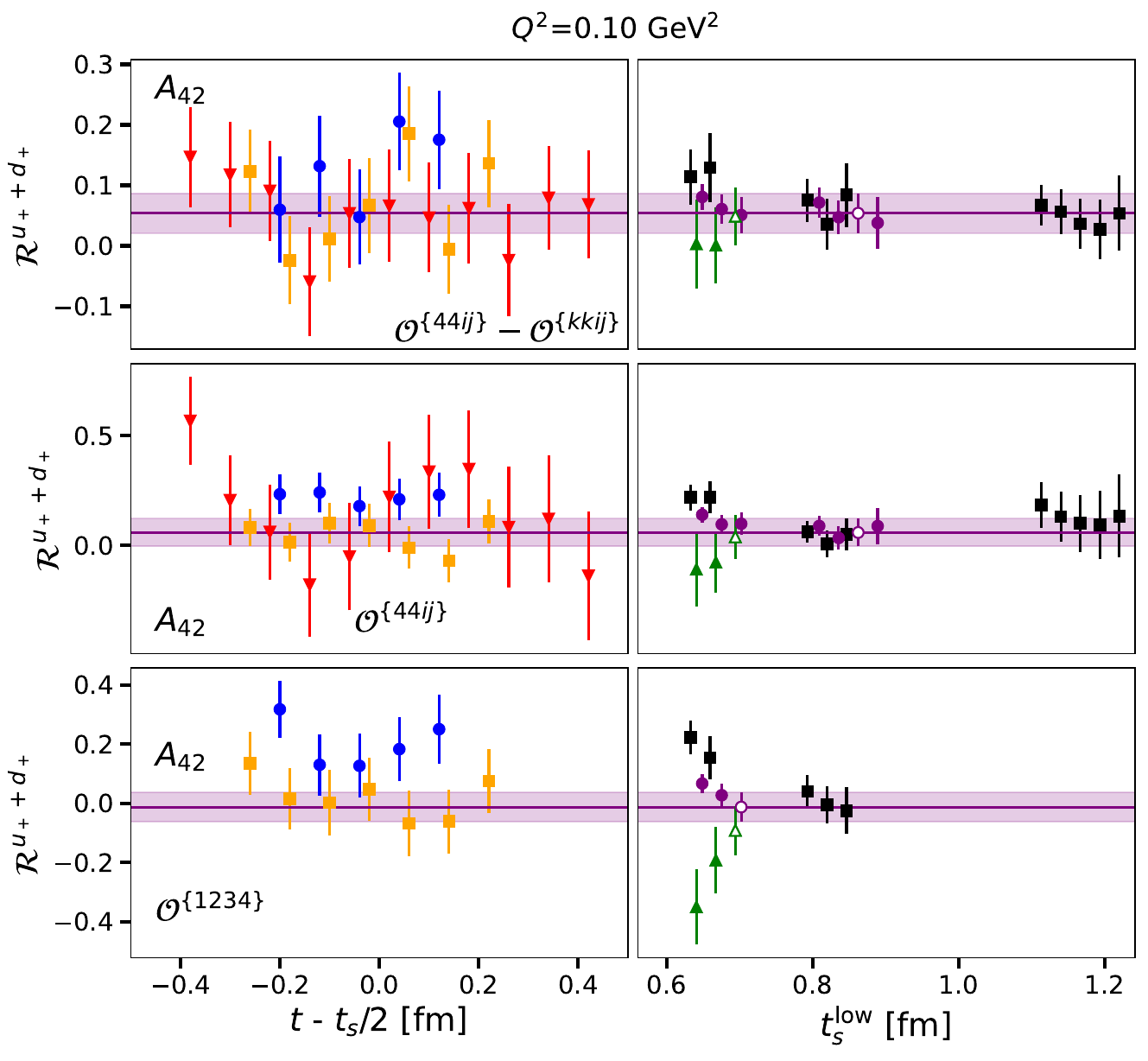}\hfill
    \caption{Same as \cref{fig:ThreeD GFFs}, but for the isoscalar combination.}
    \label{fig:ThreeD GFFs isos}
\end{figure*}
The fits are done with the summation and the plateau methods using the correlated ratios of \cref{eq:Ratio}. The convergence of the summation  and the plateau methods as we vary the lowest time separation, $t_s^{\rm low}$, used in the fits is shown. We also show the convergence as we leave out more time slices from the source and sink,  $t_{\rm cut}$ in the plateau and summation fits.  We select as final value the plateau value that shows convergence and agrees with the summation method. It turns out that this is when we use $t_s^{\rm low}=8a$ ($10a$) and $t_{\rm cut }=4a$ for the two (three)-derivative operator, as this is the earliest case where all plateau fits  converge and are compatible with the converged summation method fits for all GFFs and $Q^2$ values. We estimate the systematic error as the difference between our final value and the converged summation method, which we estimate with the value at $t_s^{\rm low}=8a$ and $t_{\rm cut}=4a$ for all GFFs and $Q^2$ considered in this work.
\begin{table}[h!]
    \centering
    \begin{tabular}{c c c}
    \hline \hline
     $q$ & $\langle x^2 \rangle_{q_-}$ & $\langle x ^3 \rangle_{q_+}$ \\
     \hline
     $u-d$ & 0.0814(50) & 0.0434(79) \\
     $u+d$ & 0.1560(65) & 0.071(11) \\
     $u$ & 0.1188(57) &  0.0573(85) \\
     $d$ & 0.0363(71) &  0.0144(44) \\
    \hline \hline
  \end{tabular}
  \caption{The Mellin moments $\langle x^2 \rangle_{q_-}$ and $\langle x^3 \rangle_{q_+}$ for different flavors $q$ obtained in the forward limit. }
  \label{tab:FL values}
\end{table}

For the three-derivative operator, the evaluation of the GFFs are more tricky. As already mentioned, we have the three operators $\mathcal{O}^{\{1234\}}$, $\mathcal{O}^{\{44ij\}}$ and $\mathcal{O}^{\{44ij\}}-\mathcal{O}^{\{kkij\}}$, which we analyze separately. We show the corresponding results on the ratios in \cref{fig:IsovectorThreeDFL} and \cref{fig:IsoscalarThreeDFL} for the isovector and isoscalar, respectively. In \cref{fig:ThreeD_A40_Q2_comp}, we show a comparison for the fourth Mellin moment for the isovector and isoscalar when using the tree operators. As can be seen,  results when using the operator $\mathcal{O}^{\{1234\}}$, which has no mixing, are much more noisy than the results when using the other two operators with the same statistics. In particular, the results when using $\mathcal{O}^{\{44ij\}}$ and $\mathcal{O}^{\{44ij\}}-\mathcal{O}^{\{kkij\}}$ are compatible with results when using $\mathcal{O}^{\{1234\}}$. Thus,  this study, we will use  the two operators $\mathcal{O}^{\{44ij\}}$ and $\mathcal{O}^{\{44ij\}}-\mathcal{O}^{\{kkij\}}$, which yield more precise results. These operators  are from the same symmetry group, so they have the same mixing with the same dimensional operators and neither can be preferred over the other. Consequently, after extracting the plateau values as described, we consider the weighted average of the two fits as our final value, also shown in Fig.~\ref{fig:ThreeD_A40_Q2_comp} for various values of the momentum transfer squared $Q^2=-q^2$. 

Our final results of the forward limit analysis for $\langle x^2 \rangle$ and $\langle x^3 \rangle$ for all flavors are shown in \cref{tab:FL values}. The errors provided in the table are the statistical error and the systematic error added in quadrature. The analysis for the $u$ and $d$ Mellin moments was carried out analogously to the isovector and isoscalar and the results are shown in \cref{fig:UpdnTwoDFL} for the third moments and in \cref{fig:Updn ThreeD GFFs} for the fourth Mellin moments.

\subsection{Momentum transfer}
In the case of momentum transfer, there are more equations and more GFFs, so one must employ the weighted least squares method described in \cref{sec:extractionmatrixelements}. We  start by showing results from our analysis of the matrix element of isovector two-derivative operators in \cref{fig:TwoD GFFs} for two different $Q^2$, and in \cref{fig:TwoD GFFs isos} for the corresponding isoscalar flavor combination.

\begin{figure*}[htb]
    \centering
    \includegraphics[width=0.9\linewidth]{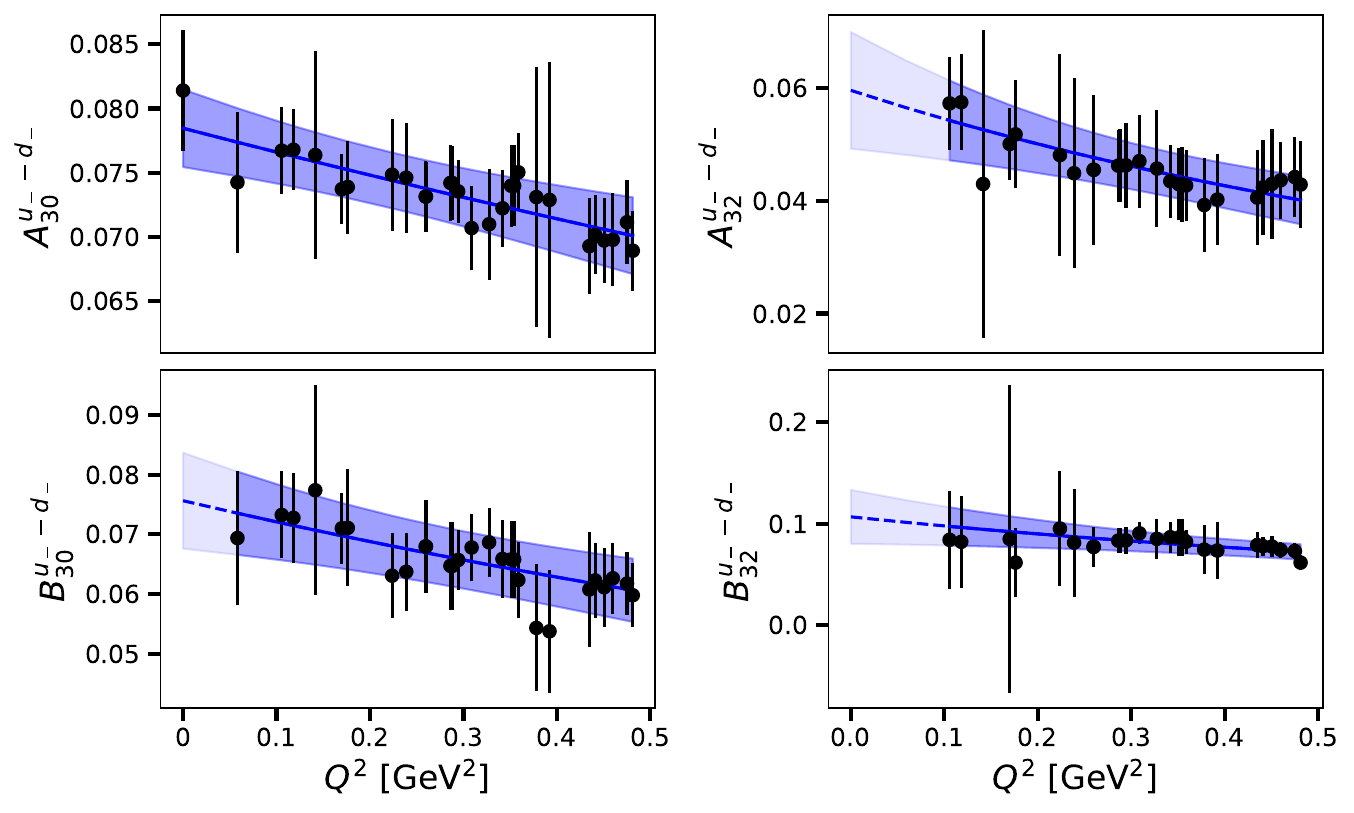}
    \includegraphics[width=0.9\linewidth]{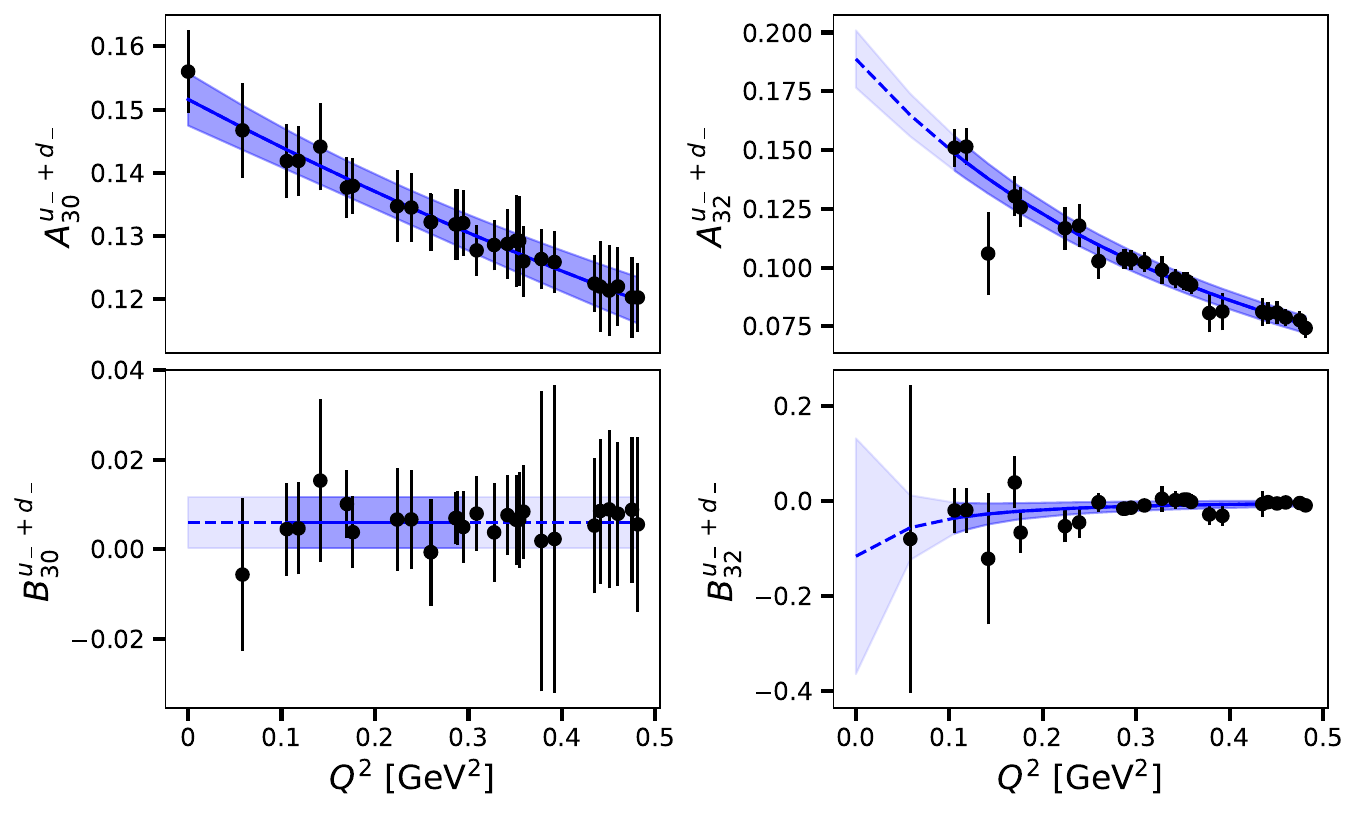}
    \caption{The isovector (top) and isoscalar (bottom) GFFs as a function of $Q^2$. The fitted region is shaded in a darker shade of blue, the extrapolation is shaded lighter.}
    \label{fig:TwoD_Gffs_dipole}
\end{figure*}

\begin{figure*}[htb]
    \centering
    \includegraphics[width=0.9\linewidth]{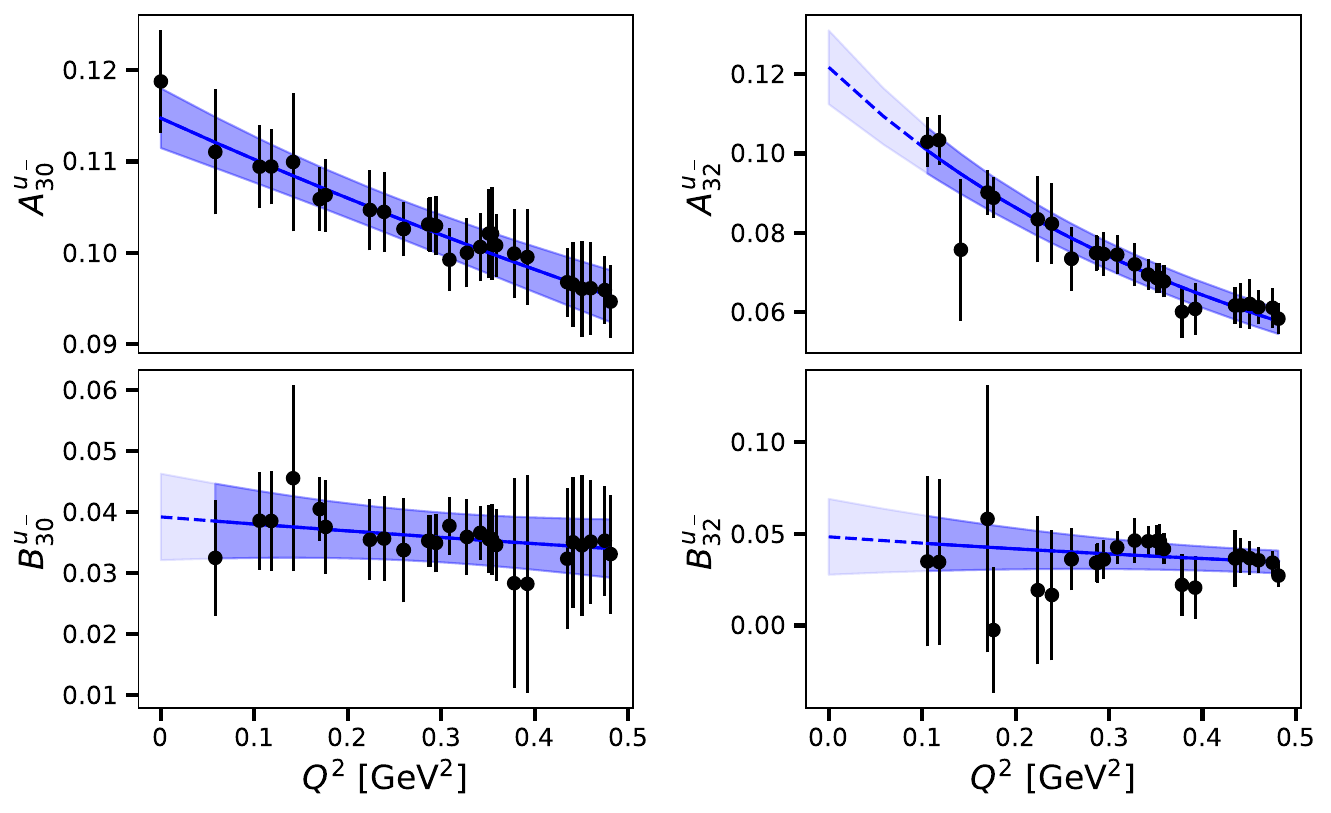}
    \includegraphics[width=0.9\linewidth]{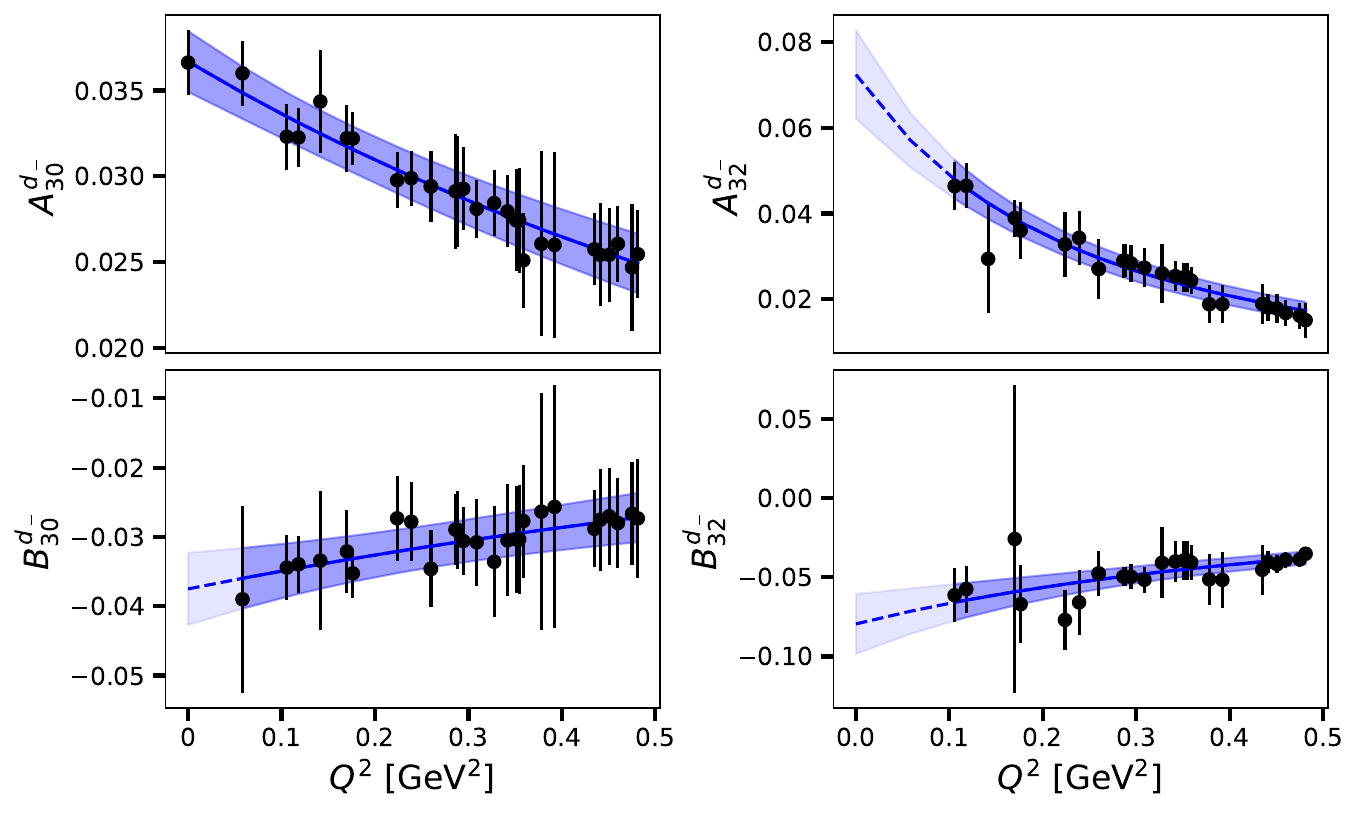}
    \caption{Same as \cref{fig:TwoD_Gffs_dipole}, but for the up (top) and down (bottom).}
    \label{fig:TwoD_Gffs_dipole_updn}
\end{figure*}

\begin{figure}[hbt]
\centering
    \includegraphics[width=0.95\linewidth]{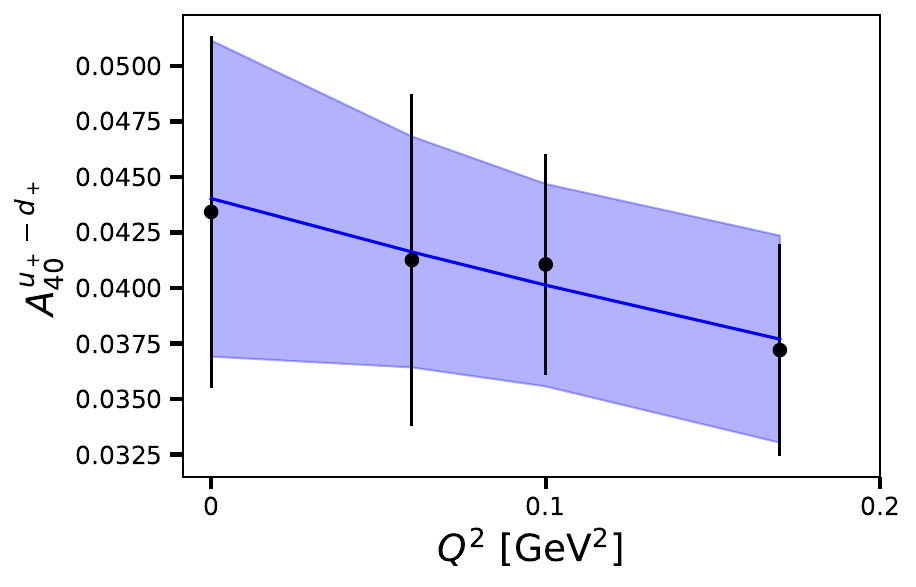}
    \includegraphics[width=0.95\linewidth]{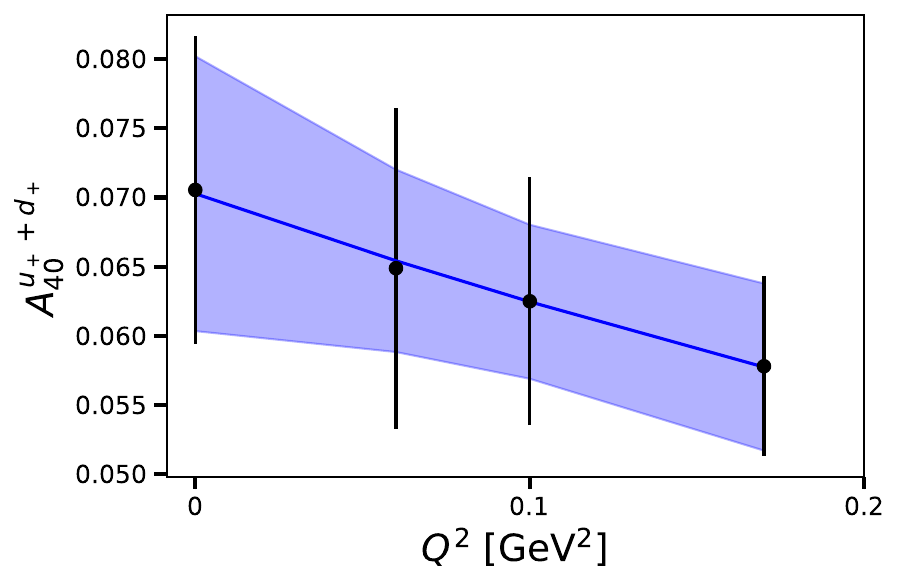}
     \caption{Resuls on the isovector $A^{u-d}_{40}(Q^2)$ as a function of $Q^2$. The points are taken from the weighted average of the two three-derivative operators $\mathcal{O}^{\{44ij\}}$ and $\mathcal{O}^{44ij}-\mathcal{O}^{kkij}$. The band show the result of the fit to the dipole form.}
    \label{fig:ThreeD_A40_dipole}
\end{figure} 

\begin{figure}[hbt]
\centering
    \includegraphics[width=0.95\linewidth]{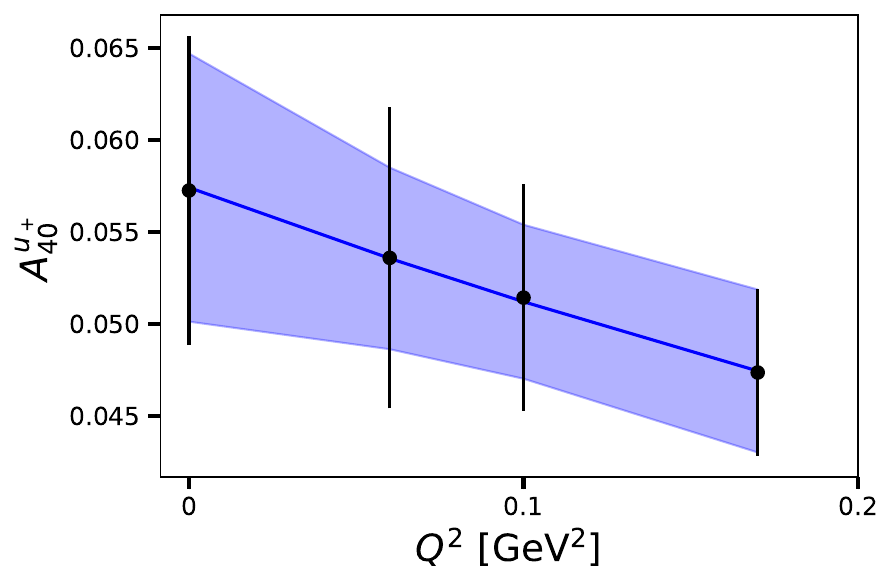}
    \includegraphics[width=0.95\linewidth]{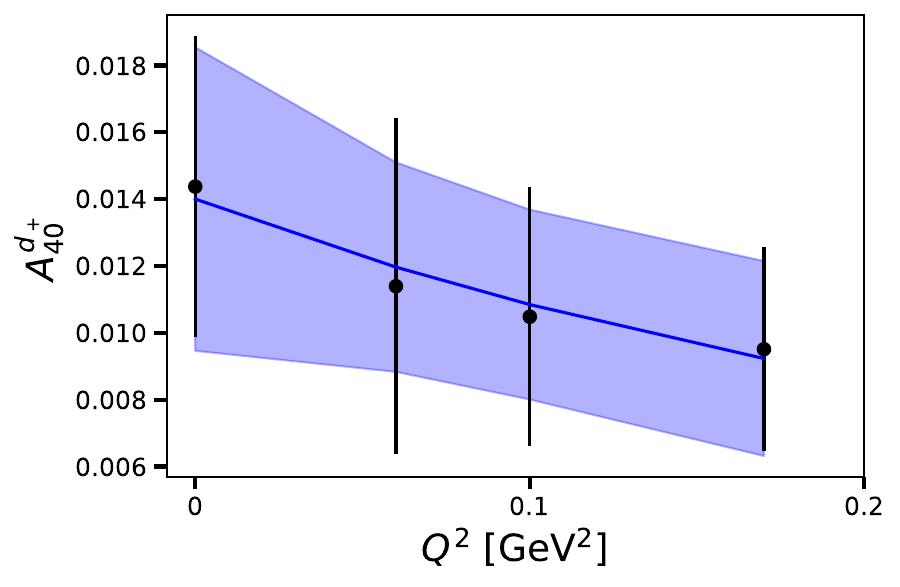}
     \caption{Same as \cref{fig:ThreeD_A40_dipole}, but for the up (top) and down (bottom).}
    \label{fig:ThreeD_A40_dipole_updn}
\end{figure}

For the matrix element of the two derivative operator the LSE has a unique solution for sufficiently many $Q^2$. However, for the three derivative operators there are often times that there are not enough linearly independent equations to solve for all GFFs. Subsequently, we cluster different equations with similar $Q^2$ together, to get enough equations. We find that we can use the values $Q^2=0.06 \GeV^2$, 0.10 $\GeV^2$ and 0.17 $\GeV^2$ for extract the GFFs, except for $C_{40}$, which is only obtainable for the last two values of $Q^2$. We perform the same analysis of the ratios as described for the case of $Q^2=0$. In \cref{fig:ThreeD GFFs}, we show results for ratio from which the isovector GFFs $A_{40}^{u_+-d_+}(Q^2)$ and $A_{42}^{u_+-d_+}(Q^2)$ at $Q^2=0.10 \GeV^2$ are extracted and and  \cref{fig:ThreeD GFFs isos}, we show the corresponding ratios  for extracting the isoscalar combination.

Having obtained the GFFs at different $Q^2$ values one can parametrize the $Q^2$-dependence using a dipole Ansatz
\begin{align}
    G(Q^2)=\frac{G(0)}{(1+\frac{Q^2}{M^2})^2}\,,
\end{align} where $G(0)$ is the GFF value at zero momentum transfer and $M$ is the dipole mass. We fit the  GFFs, after removing outliers which carry very large errors, for all $Q^2$ values  available. The errors at each $Q^2$ are given by the statistical error to which we add the systematic error. We assume that the systematic error is distributed like a Gaussian with zero mean and standard deviation being the difference between the plateau fit and summation method fit as before. We further assume no correlation between the systematic and statistical error.

The dipole Ansatz provides a good fit to all isovector GFFs as shown in \cref{fig:TwoD_Gffs_dipole}. For the isoscalar GFFs, shown in \cref{fig:TwoD_Gffs_dipole} as well, the dipole Ansantz provides a good fit for $A_{30}^{u_-+d_-}(Q^2)$, $A_{32}^{u_-+d_-}(Q^2)$ and $B_{32}^{u_-+d_-}(Q^2)$. However, for $B_{30}^{u_-+d_-}(Q^2)$ the data is roughly constant for the entire $Q^2$ region available. Thus we are fitting $B_{30}^{u_-+d_-}(Q^2)$ to a in the region between $0.1 \GeV^2 \leq Q^2 \leq 0.3 \GeV^2$, as the zero$^{\rm th}$-order term of the dipole.

For the fourth-order GFFs, as shown in the forward limit, extracting them using the operator  without mixing results in larger errors  than when using the three-derivative operators $\mathcal{O}^{\{44ij\}}$ and $\mathcal{O}^{\{44ij\}}-\mathcal{O}^{\{kkij\}}$ that have mixing. Still, the latter  two operators give consistent results  for all GFFs  compared to the one without mixing. 
 In \cref{fig:ThreeD_A40_Q2_comp}, we show the results when using the three operators for different $Q^2$ for the most precise GFFs $A_{40}(Q^2)$ for both isovector and isoscalar combinations. One can clearly see that all three operators are compatible with each other at all values of $Q^2$. So, as done in the forward limit, we take the weighted average of the results when using  $\mathcal{O}^{\{44ij\}}$ and $\mathcal{O}^{\{44ij\}}-\mathcal{O}^{\{kkij\}}$ as our final value.  The resulting dipole fits for $A_{40}(Q^2)$ of the weighted average for both isovector and isoscalar are shown in \cref{fig:ThreeD_A40_dipole}. All other GFFs are consistent with zero and are shown.

The same procedure that is used to extract the isovector and isoscalar combination are applied to determine the $u$- and $d$-quark GFFs. Their dipole fits for all the two-derivative GFFs are shown in \cref{fig:TwoD_Gffs_dipole_updn}, while the dipole fit of $A_{40}$ for the up and down are shown in \cref{fig:ThreeD_A40_dipole_updn}. The results on the dipole parameters  all GFFs fitted to a dipole for all flavors are given in \cref{tab:DipoleFit values}.

\begin{table*}[htb]
    \centering
    \begin{tabular}{c | c c | c c | c c | c c}
    \hline \hline
     & \multicolumn{2}{c}{$u-d$} & \multicolumn{2}{c}{$u+d$} & \multicolumn{2}{c}{$u$} & \multicolumn{2}{c}{$d$}\\
     & $G(0)$ & $M$ [\GeV] & $G(0)$ & $M$ [\GeV] & $G(0)$ & $M$ [\GeV] & $G(0)$ & $M$ [\GeV] \\
     \hline
     $A_{30}^{q_-}$ & 0.0785(30) & 2.88(74) & 0.1516(41) & 1.96(18) & 0.1147(33) & 2.22(26) & 0.0367(18) & 1.50(17) \\
     $A_{32}^{q_-}$ & 0.060(10) & 1.49(43) & 0.189(12) & 0.912(46) & 0.1217(93) & 1.033(76) & 0.072(10) & 0.680(65) \\
     $B_{30}^{q_-}$ & 0.0756(80) & 2.03(60) & 0.0059(57) & -       & 0.0392(71) & 2.6(2.1) & -0.0375(52) & 1.66(52) \\
     $B_{32}^{q_-}$ & 0.106(27) & 1.52(55) & -0.12(25) & 0.13(23) & 0.048(21) & 1.6(1.1) & -0.079(19) & 1.03(21) \\
     \hline
     $A_{40}^{q_+}$ & 0.0440(69) & 1.45(78) & 0.0703(95) & 1.28(56) & 0.0574(71) & 1.30(51) & 0.0140(43) & 0.86(44) \\
    \hline \hline
  \end{tabular}
  \caption{The dipole parameters extracted by fitting the GFFs to a dipole form for various flavors. In the fist row we give the GFF in the next rows from left to right, the parameters $G(0)$ and $M$ for the isovector, for the isoscalar, for the up and down. The GFF $B_{30}^{u_-+d_-}$ has been fit to a constant, so there is no mass $M$ in this fit.}
  \label{tab:DipoleFit values}
\end{table*}

\section{Comparison of results with other studies}
There have been many studies on the nucleon charges and form factors, which are the first moments of PDFs and GPDs, see Ref.~~\cite{FlavourLatticeAveragingGroupFLAG:2024oxs} for a collection of lattice results on charges.  There have also been calculations of the  second Mellin moments of PDFs and GPDs  mainly on the unpolarized moments including a full flavor decomposition. For a review on recent results on the first and second moments see Ref.~\cite{Alexandrou:2026soz}. Results on the
nucleon isovector unpolarized third PDF Mellin moment using
local operators appeared  only very recently~\cite{Taggi:2026skn}, as well as our preliminary results on the fourth Mellin moment~\cite{Alexandrou:2026soz}.   Third and fourth GFFs for the nucleon isoscalar have not been  presented before. 
\subsection{Comparison of Mellin moments of PDFs}
\begin{figure}[htb]
\centering
    \includegraphics[width=0.95\linewidth]{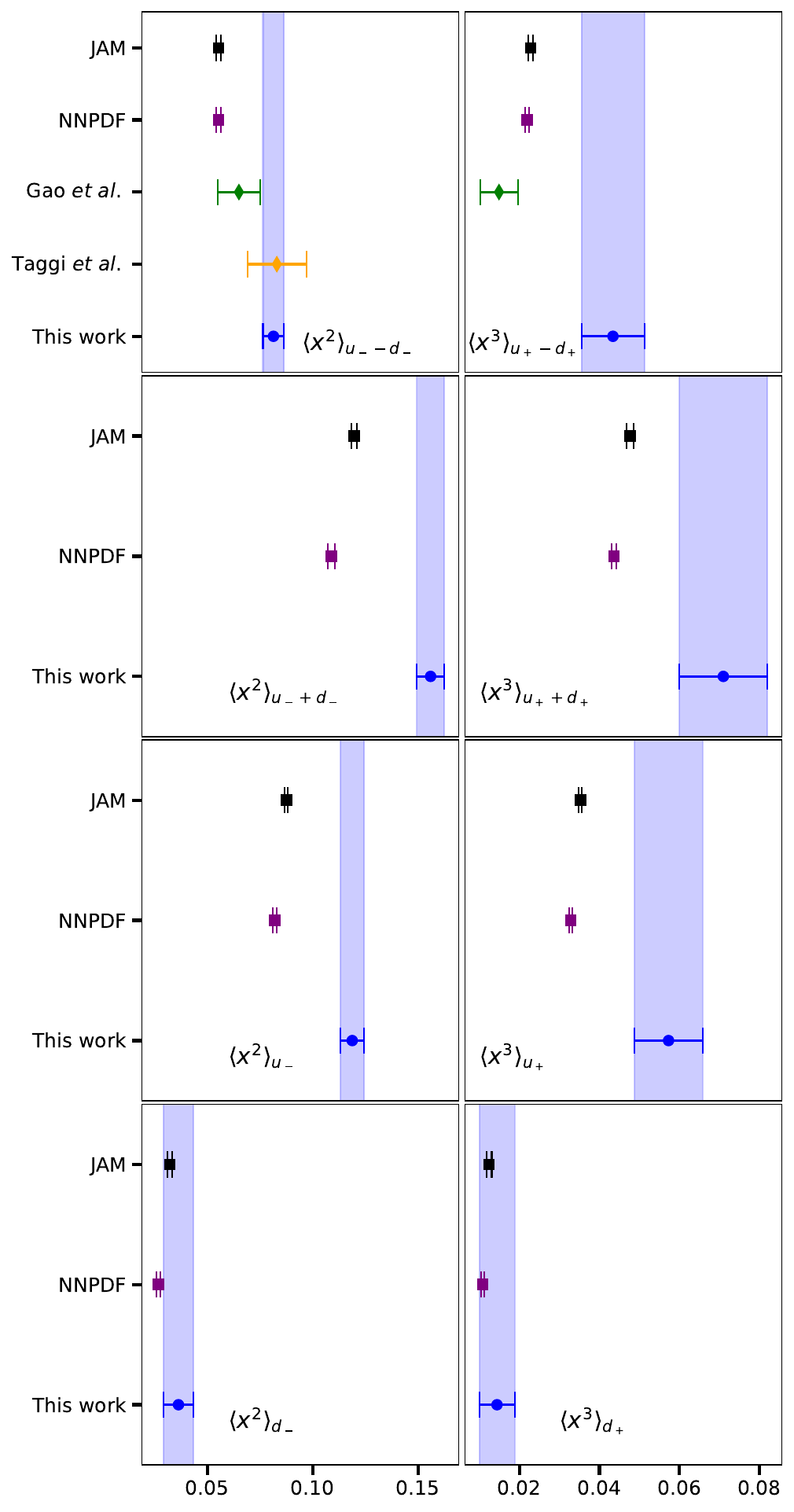}
     \caption{Results on the third (left) and fourth (right) nucleon moment from lattice QCD calculations and phenomenology. JAM~\cite{Cocuzza:2022jye} (black circle) and NNPDF~\cite{NNPDF:2021njg} (purple circle) are phenomenological studies, while Gao \etal~\cite{Gao:2022uhg} (green) and Taggi \etal~\cite{Taggi:2026skn} (orange) are lattice QCD studies, the former using non-local operators and the latter using local ones as in this work.}
    \label{fig:Comp_literature}
\end{figure}

Within the lattice community, there has been recent progress in extracting Mellin moments using non-local operators either in the quasi- of pseudo-distribution approaches~\cite{Ji:2013dva,Ji:2014gla,Radyushkin:2017cyf,Orginos:2017kos,Gao:2022uhg,Alexandrou:2026soz}. This allows us to compare our results for the isovector Mellin moments with those extracted using matrix elements of non-local operators. All the results up to date for the higher moments are done  for the isovector case. For the isoscalar and the $u$- and $d$-quark distributions we can only compare with phenomenological studies, such those from JAM~\cite{Cocuzza:2022jye} and NNPDF~\cite{NNPDF:2021njg}. We provide a comparison of our results on the isovector Mellin moments with those extracted using the non-local operator~\cite{Gao:2022uhg}  in \cref{fig:Comp_literature}.  In their work Gao \etal used a $64^4$  ensemble with highly-improved staggered quarks simulated with  physical pion mass  with a lattice spacing of $a=0.076\fm$. Our value for the third moment is compatible with theirs but larger that the the values found by JAM and NNPDF. Similarly, our fourth moment is larger by about two standard deviations. The  other  recent calculation of the third Mellin moment using local operators is by Taggi \etal~\cite{Taggi:2026skn}, where they used a $48^4$ and a $64^4$ clover fermion ensemble with lattice spacings of $0.1163(4)\fm$ and $0.0926(6)\fm$, respectively, to estimate the continuum limit. Like in our work, both their ensembles are at the physical pion mass, but  their lattices are courser than ours cf. \cref{tab:gauge-ensemble}.    Our value is compatible with what they estimate in the continuum limit indicating that  lattice cutoff effects are small compared to statistical errors. Since both the third and the fourth Mellin moments  are larger than  phenomenological results, the unpolarized isovector PDF that we will determine will have a larger support at large $x$ values, see next section.

For the isoscalar and up and down quark flavors only  phenomenological studies are available.   We remind that we only include the connected contributions to the isoscalar, and subsequently, also to the up and down quark Mellin moments. Neglecting these contributions that are expected to  play a more important role at low $x$ and, thus,  to be more suppressed for these higher moments than for the first and second moments, we compare in \cref{fig:Comp_literature} our results on the isoscalar and $u$ and $d$ Mellin moments with phenomenological results. We find that our results for the isoscalar and $u$ Mellin moments are larger than those from the two phenomenological analyses. For the $d$ Mellin moments our values are much closer to the those from the two phenomenological analyses.

\subsection{Reconstruction of PDFs\label{sec:Reconstruction}}
Having determined the PDF Mellin moments, we can attempt to reconstruct the PDFs  using the standard Ansatz for the nucleon PDF
\begin{align}
    q(x) = N x^\alpha (1-x)^\beta \,,
    \label{eq:pdfAnsatz}
\end{align} with fit parameters $\alpha$ and $\beta$. Integrating this Ansatz between 0 and 1 gives for the moments
\begin{align}
    \langle x^n \rangle &= \int_0^1\,{\rm d}x \, x^nq(x) \nonumber \\
    &= \int_0^1\,{\rm d}x \, x^nN x^\alpha (1-x)^\beta = N B(\alpha+n+1,\beta+1)\,,
\end{align} with the beta function $B(x,y)$. If we consider the valence quark distribution $q_v=q_-$ the normalization constant $N$ can be fixed by valence quark sum rules. However, there is a nuance concerning the anti-quark distributions:
\begin{figure}[htb]
\centering
    \includegraphics[width=0.95\linewidth]{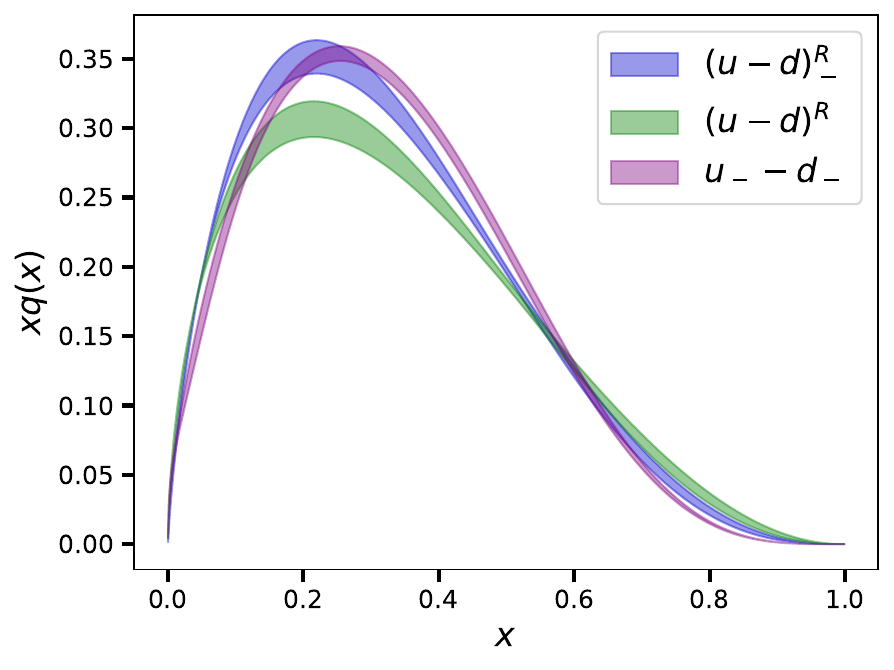}
     \caption{A comparison of isovector PDFs. The blue band $(u-d)_-^R$ shows the reconstruction with \cref{eq:pdfAnsatz} of the moments $\langle x \rangle_{u_--d_-}$, $\langle x^2 \rangle_{u_--d_-}$ and $\langle x^3 \rangle_{u_--d_-}$. The green band shows the reconstruction of the moments $\langle x \rangle_{u_+-d_+}$, $\langle x^2 \rangle_{u_--d_-}$ and $\langle x^3 \rangle_{u_-+d_+}$. The moments were taken from JAM~\cite{Cocuzza:2022jye}. The purple band shows the $u_--d_-$ PDF from JAM for comparison.}
    \label{fig:Reconstructing_JAM}
\end{figure}
\begin{figure}[htb]
\centering
    \includegraphics[width=0.95\linewidth]{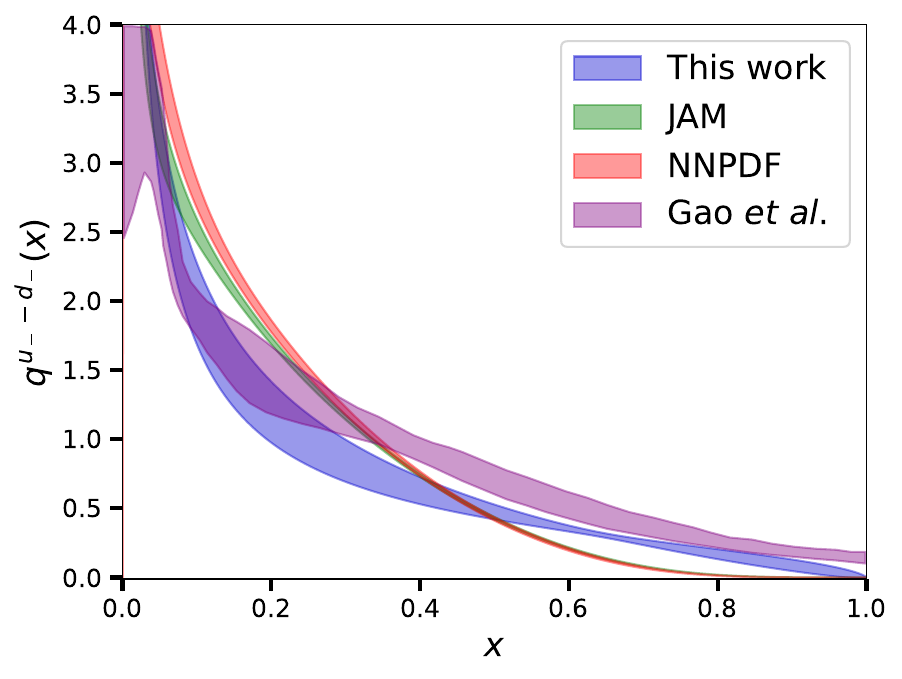}
     \caption{Results on the reconstructed nucleon isovector valence quark PDF (blue) compared to JAM (green) and NNPDF (red) data, and to the result of the lattice group by Gao (purple).}
    \label{fig:Reconstruction}
\end{figure}
For Mellin moments $\langle x^n \rangle$ with even $n$, the anti-quark distribution gets subtracted, while it is added for odd $n$, as explained above, see \cref{eq:defMellinmoment}. Therefore, one can in general not mix odd and even moments together, and they must be studied separately. Subsequently, one must use at least up to the sixth Mellin moment for the odd-$n$ terms and the fifth for the even-$n$ terms. This would require up to five derivatives, which one cannot do using local operators, unless new approaches like those proposed recently prove efficient. We note that higher moments, besides being more noisy, they are smaller and, thus, have less and less impact on the reconstruction, as they can only describe the large $x$ ($x \approx 1$) behavior, because they are  less sensitive to the small and intermediate $x$ values. 

Instead of going to higher order moments, we are instead supplementing our results with phenomenological ones. Indeed, for odd-$n$ moments one finds
\begin{align}
    \langle x^{n}\rangle_{q_-}=\langle x^{n}\rangle_{q_+}-2\langle x^{n}\rangle_{\bar{q}}\,,
\end{align} which means that by subtracting the double of the anti-quark moment one can obtain $\langle x^n \rangle_{q_-}$ also for odd $n$. This is not possible on the lattice directly, where quark and anti-quark contributions are always mixed together. Therefore we take the anti-quark contribution from JAM~\cite{Cocuzza:2022jye} to obtain the correct distribution. 

Let us now consider the isovector: While Regge theory suggests that $\int_0^1\,{\rm d}x \, [\bar{d}(x) -\bar{u}(x)]$ is finite, it does not have to be zero. This relation is encoded in the Gottfried sum rule, which is broken experimentally as is also supported by theory~\cite{Kataev:2003xp}. The effects of including the anti-quark moments is noticeable as shown in \cref{fig:Reconstructing_JAM}. The reconstruction with the moments $\langle x \rangle_{u_--d_-}$, $\langle x^2 \rangle_{u_--d_-}$ and $\langle x^3 \rangle_{u_--d_-}$ is similar to the JAM PDF. A perfect match to the JAM PDF is not expected, as we ignore correlations between the moments, and JAM uses a different parametrization of the PDF than \cref{eq:pdfAnsatz}. However, that reconstruction is much closer to JAM than the reconstruction with the moments $\langle x \rangle_{u_+-d_+}$, $\langle x^2 \rangle_{u_--d_-}$ and $\langle x^3 \rangle_{u_-+d_+}$. This demonstrates that even small anti-quark distributions are important. For the flavors $q=u$, $d$ and $u+d$ the anti-quark distribution is much more significant, as the integral $\int_0^1 \,{\rm d}x \, q(x)$  diverges for all partons $q$~\cite{Bali:2018zgl}. Thus, we will not discuss further these flavors, as the anti-quark distributions are not just corrections, but similar in size to the actual moments.

Now we have to calculate the moments $\langle x^n \rangle_{u_--d_-}$. The second moment $\langle x \rangle_{u_+-d_+}$ has already been calculated in Ref.~\cite{Alexandrou:2019ali}. However,  more statistics were recently analyzed and the renormalization function  have been updated. Thus we use updated value  $\langle x \rangle_{u_+-d_+}=0.168(13)$. We refer to~\cite{Yansnewpaper} for more details. The moments $\langle x^2 \rangle_{u_--d_-}$ and $\langle x^3 \rangle_{u_+-d_+}$ are taken from \cref{tab:FL values}, and the anti-quark corrections $-2\langle x \rangle_{\bar{u}-\bar{d}}=0.0152(20)$ and $-2\langle x^3 \rangle_{\bar{u}-\bar{d}}=0.00049(45)$ are from JAM~\cite{Cocuzza:2022jye}. In the end we are fitting the moments
\begin{align}
    \langle x \rangle_{u_--d_-} &=0.183(13) \nonumber\\
    \langle x^2 \rangle_{u_--d_-}&=0.0814(51) \nonumber\\ 
    \langle x^3 \rangle_{u_--d_-}&=0.0439(78)\,,
\end{align} to the Ansatz \cref{eq:pdfAnsatz}, and we find 
\begin{align}
    \alpha= -0.59(12) \,, \quad \beta=0.83(40)\,.
\end{align} The resulting fit is shown in \cref{fig:Reconstruction} and compared to JAM, NNPDF and the result from Gao \etal~\cite{Gao:2022uhg}. As we have larger values for the  third and fourth Mellin moments, it is not surprising that our PDF is larger at larger $x$ values  compared to the phenomenologically determined PDF. This is also reflected in the smaller value of $\beta$, which determines the large $x$ behavior. For most $x$ our curve is also compatible with the result from Gao, but with better behavior at the edges.

\section{Conclusions}
In this work, we present the  first  calculation of  nucleon Mellin moments of the unpolarized PDF up to fourth order within lattice QCD. Furthermore, we calculate the $Q^2$-dependence of the GFFs. Our results are renormalized nonperturbatively in the RI$'$-MOM scheme and matched perturbatively to the $\overline{\rm MS}$ scheme at the scale $\bar{\mu} = 2$ \GeV. For the third moments of GPDs we perform dipole fits to most GFFs and a constant fit to $B_{32}^{u_-+d_-}(Q^2)$ and determine the forward limit. For the fourth GPD moments, we find that only  $A_{40}(Q^2)$ is not consistent with zero, while the other four are consistent with zero within statistical precision. We estimate the systematic error by the difference between the plateau method fit and the summation method result for each GFF and $Q^2$ and add it to the statistical error in quadrature. Furthermore we combine phenomenological results for the anti-quark PDFs with our moments to reconstruct the isovector valence PDF $u_--d_-$, and compare it with both phenomenological results and a lattice study.

A comparison of the isovector Mellin moments shows that we are compatible with the recent result from Taggi \etal~\cite{Taggi:2026skn} with our value being more precise. Compared to phenomenological studies, our values are larger. Since statistical errors are larger for higher Mellin moments, a thorough analysis of lattice systematics would need larger statistics as compared to what we could currently achieve given the available computational resources. New techniques, like using Wilson flow~\cite{Francis:2025pgf,Francis:2025rya}, is a promising way for the extraction  of larger Mellin moments, once proved effective.  Having these Mellin moments using an established lattice QCD methodology will serve as a benchmark for the Wilson flow-based approach using the same gauge ensemble.

\acknowledgments 
We would like to thank all members of ETMC for the most enjoyable collaboration.
This research is supported by the European Union’s HORIZON MSCA Doctoral Networks programme, under Grant Agreement No.\ 101072344,  AQTIVATE (Advanced computing, QuanTum algorIthms and data-driVen Approaches for science, Technology and Engineering). The project is implemented under the programme of social cohesion “3D-nucleon" (EXCELLENCE/0421/0043), "IMAGE-N" (EXCELLENCE/0524/0459), "MuonHVP" (EXCELLENCE/0524/0017), "PulseQCD"
(EXCELLENCE/0524/0269), "StrongILA" (EXCELLENCE/0524/0001), "partonWF" (VISIONERC/0525/0010)”, and "$\varphi$BSM" (CULTURE/AWARD-YR/0524B/0002) co-funded by the European Union, through Research and Innovation Foundation. M.~C. acknowledges financial support by the U.S. Department of Energy, Office of Nuclear Physics,  under Grant No.\ DE-SC0025218.

An award of computer time was provided by the U.S. Department of Energy’s (DOE) Innovative and Novel Computational Impact on Theory and Experiment (INCITE) Program. This research used supporting resources at the Argonne and the Oak Ridge Leadership Computing Facilities. The Argonne Leadership Computing Facility at Argonne National Laboratory is supported by the Office of Science of the U.S. DOE under Contract No. DE-AC02-06CH11357. The Oak Ridge Leadership Computing Facility at the Oak Ridge National Laboratory is supported by the Office of Science of the U.S. DOE under Contract No. DE-AC05-00OR22725.%Polaris https://docs.alcf.anl.gov/policies/alcf-acknowledgement-policy/#incite-alcf-acknowledgement
The authors gratefully acknowledge the Gauss Centre for Supercomputing e.V. (www.gauss-centre.eu) for funding this project by providing computing time through the John von Neumann Institute for Computing (NIC) on the GCS Supercomputer JUWELS, JUWELS Booster~\cite{JUWELS} and JUPITER Booster 
at Jülich Supercomputing Centre (JSC).%Juwels https://www.fz-juelich.de/en/jsc/systems/supercomputers/apply-for-computing-time/gcs-nic?activeAccordion=ba7248cd-006b-4b7e-963e-6cef5c218522
This work was supported by a grant from the Swiss National Supercomputing Centre (CSCS) under project IDs ch15 and lp139 on Alps.%alps https://www.cscs.ch/user-lab/code-of-conduct

\bibliographystyle{apsrev4-2}
\bibliography{literature}

\end{document}